%% file: bare_jrnl_new_sample4.tex
\documentclass[lettersize,journal]{IEEEtran}
\usepackage{amsmath,amsfonts}
\usepackage{algorithm}
\usepackage{array}
\usepackage[caption=false,font=normalsize,labelfont=sf,textfont=sf]{subfig}
\usepackage{textcomp}
\usepackage{stfloats}
\usepackage{url}
\usepackage{verbatim}
\usepackage{graphicx}
\usepackage{cite}

\usepackage{tikz}
\usepackage{amsmath}
\usepackage{booktabs}
\usepackage{makecell}
\usepackage{algorithm}
\usepackage{algpseudocode}
\usepackage{graphicx}
\usepackage{subcaption}
\usepackage{xcolor}
\usepackage{textcomp}
\usepackage{xspace}
\usepackage{tabularx} % 导言区添加
\usepackage{multirow}
\usepackage{pifont}
\usepackage{amssymb}   
\usepackage{pgfplots}
\pgfplotsset{compat=1.18}
\usepgfplotslibrary{groupplots}
\usepgfplotslibrary{polar}
\usepackage{subcaption}
\usepackage{enumitem}
\usetikzlibrary{patterns}
\usepackage{orcidlink}

\input{micros.tex}

\begin{document}

\title{FFSlim: An Efficient and Lightweight Format for Multi-modal Data Storage and Retrieval}

\author{Long Yang\orcidlink{0009-0001-6566-6917}, Yu Mao\orcidlink{0000-0001-9803-4927}, Yuchen Shao\orcidlink{0009-0009-0414-7521}, Yumiao Zhao\orcidlink{0000-0001-7803-2887}, Yaqi Li\orcidlink{0009-0007-6040-2609}, Xuan Liu\orcidlink{0009-0006-0093-0912}, Xiaolong Shen\orcidlink{0009-0005-1331-8936}, Tao Yu\orcidlink{0009-0003-1996-0319}, Gezi Li\orcidlink{0009-0005-6766-7971}, Jing Wang\orcidlink{0000-0001-7260-0521}, Chengcheng Wan\orcidlink{0000-0001-9162-9688}, and Liang Shi\orcidlink{0000-0002-9977-529X}, Member, IEEE}



% % column for its text to clear the IEEEpubid mark.

\maketitle

\begin{abstract}
With the rapid expansion of large-scale media–text corpora, multi-modal datasets increasingly require efficient storage and retrieval. 
Existing formats such as \textit{Files}, \textit{TDP}, and \textit{FFRecord} work adequately for uni-modal data but expose fundamental limitations in multi-modal settings, including storage redundancy, massive small-file overheads, cache-unfriendly layouts, and heavy index structures. 
These issues jointly inflate storage and memory usage and make I/O the dominant bottleneck in real training workloads.
We present \textit{FFSlim}, a lightweight format for storing and retrieving multi-modal data. 
FFSlim improves storage efficiency and loading throughput through three components: a unified file format that removes media duplication and avoids small-file proliferation; 
an adaptive retrieval mechanism that enables low-overhead pair-level access and accelerates repeated media loading; 
and a redundancy detection and aggregation module that converts existing datasets into the FFSlim layout.
The experimental results demonstrate that \textit{FFSlim} achieves $2.07\times$ and $8.26\times$ higher data-loading and write throughput on average than the strongest baseline, with minimal storage and index overhead. 
Consequently, these underlying I/O accelerations enable \textit{FFSlim} to reduce end-to-end training time by $5.36\%$--$14.18\%$ across seven diverse multi-modal models.
\end{abstract}


\section{Introduction}
% Context & Problem Statement
Multi-modal models have demonstrated exceptional performance across a wide spectrum of artificial intelligence tasks~\cite{liu2023visual} and are increasingly deployed in diverse real-world application scenarios~\cite{xie2024osworld}. 
However, recent profiling studies reveal that data I/O has emerged as a severe bottleneck in multi-modal training~\cite{ImPACT}. 
This bottleneck fundamentally stems from a widening hardware performance gap: 
AI accelerators (\eg, GPUs, ASICs) have advanced at a much faster pace than underlying storage devices~\cite{zhao2023silod}. 
This physical I/O disparity is further compounded by modern training pipelines, which increasingly rely on massive, heterogeneous multi-modal datasets~\cite{MLDataForge}.
For instance, WebLI~\cite{wang2025scaling100b}, the largest open-source image-text dataset, contains over 100 billion pairs. 
Furthermore, training frameworks rigorously reshuffle these massive datasets at the beginning of every epoch to ensure statistical randomness, severely disrupting spatial locality and exacerbating I/O inefficiency~\cite{chen2023icache}.

\begin{figure}[t]
\centerline{\includegraphics[width=1\linewidth]{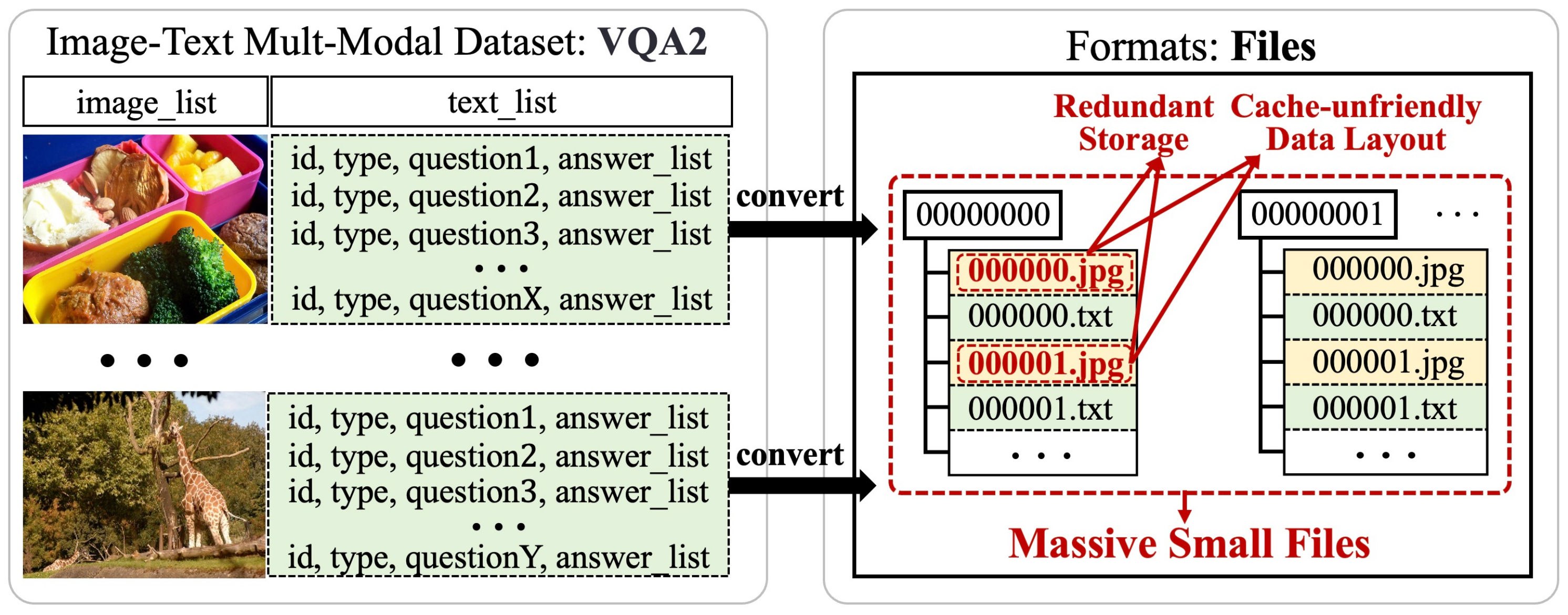}}
\caption{A storage example of an image-text multi-modal dataset. \footnotesize{Naive multi-modal data organization suffers from several fundamental limitations.}}
\label{example4formats}
\end{figure}

Several methods have been proposed to address this problem.
The \textit{Files} format~\cite{radford2021learning} stores each sample’s media and text as separate files organized by directory hierarchy and naming rules, as shown in Figure~\ref{example4formats}. 
During training, indices are computed on the fly to load samples when needed. 
This design is simple and works reasonably well for uni-modal datasets. 
However, on multi-modal datasets, it suffers from two major issues: 
high storage usage and poor I/O performance. 
These limitations come from storage redundancy, a very large number of small files, and a data layout that does not work well with common caching mechanisms.
Recent work explores two directions.
\textit{TDP}~\cite{alayrac2022flamingo} separates modalities to reduce redundancy, cut down small files, and provide a more cache-friendly layout. 
This improves storage usage and I/O efficiency, but it still leaves many small files unresolved and introduces heavy indexing with high memory cost.
\textit{FFRecord}~\cite{ffrecord} packs the dataset into a single large file, which removes the small-file issue and improves I/O performance. 
However, it ignores storage redundancy and keeps a cache-unfriendly layout, resulting in high storage usage and limited loading speed.

Multi-modal data naturally follow a \textit{one-to-many} structure rather than a one-to-one pattern. 
This structure is ubiquitous across image–text, video–text, and audio–text datasets (see Table~\ref{tab:dataset}), where a single media object is typically paired with multiple text descriptions, captions, questions, or narrations. 
This one-to-many structure arises from several inherent properties of multi-modal data. 
Real-world media are inherently multi-faceted, and no single text description can fully capture their semantic richness. 
This drives practitioners to collect multiple linguistic views, a process that is far more efficient than acquiring new media because text annotations are inexpensive and scalable. 
These diverse aligned texts, in turn, provide richer cross-modal supervision, enabling models to approximate human-like multi-modal cognition and learn more robust associations.
However, existing storage systems such as TDP and FFRecord do not incorporate this structural property into their design. 
They continue to manage data at the media–text–pair granularity, implicitly assuming a one-to-one structure. 
As a result, they suffer significant inefficiencies and performance degradation on modern multi-modal workloads that rely heavily on one-to-many relationships.

\begin{table}[t]
\centering
\caption{Statistics of representative multi-modal datasets. \footnotesize{In most multi-modal datasets, a single media object is associated with multiple text descriptions, and the media objects are typically far larger than the text.}}

\label{tab:dataset}
\resizebox{\columnwidth}{!}{%
\begin{tabular}{l!{\vrule}c|cccccc}
\toprule
\multirow{2}{*}{\centering \textbf{Dataset}}  
& \multirow{2}{*}{\centering \textbf{Modality}} 
& \multicolumn{2}{c}{\textbf{Quantity}} & \textbf{Quantity} & \multicolumn{2}{c}{\textbf{Avg Size}} & \textbf{Avg Size} \\
\cmidrule(lr){3-4} \cmidrule(lr){6-7}
 &  & \textbf{Media} & \textbf{Text} & \textbf{Avg Ratio} & \textbf{Media} & \textbf{Text} & \textbf{Ratio} \\
\midrule
Flickr30k~\cite{young2014image} & image-text & 31K & 155K & 1:5.0 & 136KB & 63B & 2207:1 \\
F30-E~\cite{plummer2017flickr30k} & image-text & 31K & 155K & 1:5.0 & 136KB & 330B & 421:1 \\
MSCOCO~\cite{lin2014microsoft} & image-text & 118K & 590K & 1:5.0 & 159KB & 52B & 3140:1 \\
VQA2~\cite{goyal2017making} & image-text & 82K & 443K & 1:5.4 & 153KB & 826B & 190:1 \\
GQA~\cite{hudson2019gqa} & image-text & 69K & 943K & 1:13.6 & 138KB & 125B & 1133:1 \\
MSRVTT~\cite{xu2016msrvtt} & video-text & 10K & 200K & 1:20.0 & 656KB & 48B & 14003:1 \\
Clotho~\cite{drossos2020clotho} & audio-text & 20K & 100K & 1:5.0 & 1933KB & 63B & 31415:1 \\
\bottomrule
\end{tabular}%
}
\end{table} 

To address these issues, we first conduct a comprehensive evaluation of \textit{Files}, \textit{TDP}, and \textit{FFRecord} on 7 representative multi-modal datasets and identify a common root cause. 
All three formats assume a one-to-one data structure, while real multi-modal datasets typically follow a one-to-many structure. 
This mismatch leads to redundant storage, excessive small files, heavy indexing overhead, and poor cache behavior.

Based on these observations, we design \textit{FFSlim}, a lightweight and efficient storage format that minimizes storage and memory usage while improving retrieval and caching performance.
It manages data at the \textit{Media2Texts} granularity, storing each media once together with all its associated texts. 
This design removes cross-modal redundancy, avoids repeated storage, and enables cache reuse of the media across samples within an epoch. 
Retrieval and caching are also performed at the \textit{Media2Texts} level. 
When a media-text pair is requested, the full object is loaded, the target pair is returned, and the remaining pairs are cached for later access. 
Since media are much larger than texts, the small amount of extra reading has negligible cost, and caching brings consistent performance gains.
\textit{FFSlim} further organizes the entire dataset in a single large file to eliminate the massive number of small files and adopts an adaptive sample-level indexing scheme that selects simple and efficient index structures based on dataset characteristics. 
This avoids the overhead of the complex, one-size-fits-all indexing used in prior formats. 
In addition, \textit{FFSlim} supports one-click conversion from existing formats.

Through these designs, \textit{FFSlim} surpass TDP and FFRecord on four aspects: storage redundancy, 
large number of small files, 
cache-unfriendly data layouts, and expensive data indexing. 
It provides a high-performance multi-modal storage format that has the potential to serve as a new community standard.

We rigorously evaluate \textit{FFSlim} against three widely used storage baselines---\textit{Files}, \textit{FFRecord}, and \textit{TDP}---across several diverse media--text multi-modal datasets. 
Extensive micro-benchmarks demonstrate that \textit{FFSlim} significantly outperforms existing formats. 
Specifically, it achieves an average speedup of $2.07\times$ in data-loading throughput and $8.26\times$ in write throughput. 
Crucially, \textit{FFSlim} delivers these substantial I/O improvements while strictly maintaining the lowest storage overhead and a minimal index memory footprint. 
Furthermore, to quantify the macroscopic benefits, we conduct comprehensive evaluations across seven diverse multi-modal models. 
By effectively eliminating I/O bottlenecks, \textit{FFSlim} successfully translates its micro-level data access efficiency into tangible end-to-end accelerations, reducing the total training time by $5.36\%$--$14.18\%$ compared to the best-performing baseline.

In summary, this paper offers the following contributions:
\begin{itemize}
    \setlength{\itemsep}{2pt} 
    \setlength{\parskip}{0pt}
    \item \textbf{Comprehensive analysis of existing formats.}  
    We evaluate 3 existing data formats across 7 multi-modal datasets and reveal that they all suffer from the same root cause: assuming a one-to-one data structure, while real-world multi-modal datasets follow a one-to-many pattern.
    
    \item \textbf{An efficient and lightweight storage format.}  
    We propose \textit{FFSlim}, which stores, manages and retrieves multi-modal data at the \textit{Media2Texts} granularity, and employs an adaptive lightweight indexing scheme tailored to dataset characteristics to enable efficient, sample-level (media–text pair) random access.

    \item \textbf{Substantial Performance and Resource Efficiency Gains.} 
    Extensive evaluation demonstrates that \textit{FFSlim} significantly enhances both data loading and writing throughput while strictly minimizing storage and index memory overhead.
\end{itemize}

\section{Existing Storage Formats and Their Structural Mismatch}
\label{sec3}

In this section, we examine three widely used multi-modal data formats, namely \textit{Files}~\cite{radford2021learning}, \textit{Table Data Path (TDP)}~\cite{alayrac2022flamingo}, and \textit{FFRecord}~\cite{ffrecord}. They are representative of current practice in the open-source community and industrial systems. We analyze their design assumptions and study their behavior across seven representative datasets. 

\subsection{Overview of Multi-modal Datasets}
Multi-modal datasets, which contain data from multiple modalities such as images, text, audio, and video, have become essential for advancing AI applications~\cite{MM-LLMs}. 
They offer a comprehensive understanding of real-world scenarios by capturing complementary information from different data sources~\cite{qiu2024mmsum}. 
They are particularly valuable in fields like computer vision, natural language processing, and speech recognition~\cite{mei2024wavcaps}. 

% 聚焦图文多模态数据集详细介绍并简单引入Motivation
For image-text datasets, early ones, such as Flickr30k~\cite{young2014image} and MSCOCO~\cite{lin2014microsoft}, enabled image description and grounding. Later, VQA~\cite{goyal2017making} and GQA~\cite{hudson2019gqa} extended them to visual reasoning. More recent efforts scale up web-harvested image-text corpora, including Conceptual12M~\cite{changpinyo2021conceptual} and the latest 100B-scale collections~\cite{wang2025scaling100b}.  
For video-text datasets, MSR-VTT~\cite{xu2016msrvtt} and HowTo100M~\cite{miech2019howto100m} established large-scale training corpora, followed by InternVid~\cite{wang2024internvid} and OpenVid-1M~\cite{nan2025openvid}, which push toward web-scale video-text pretraining and generation.  
For audio-text datasets, Clotho~\cite{drossos2020clotho} and WavCaps~\cite{mei2024wavcaps} are audio captioning benchmarks, while recent large-scale resources such as LAION-Audio~\cite{wu2023large} and AudioSetCaps~\cite{bai2025audiosetcaps} further enrich multi-modal audio-language research.

\subsection{Existing methods}
\label{sec3:formats}

\begin{figure}[t]
\centerline{\includegraphics[width=1\linewidth]{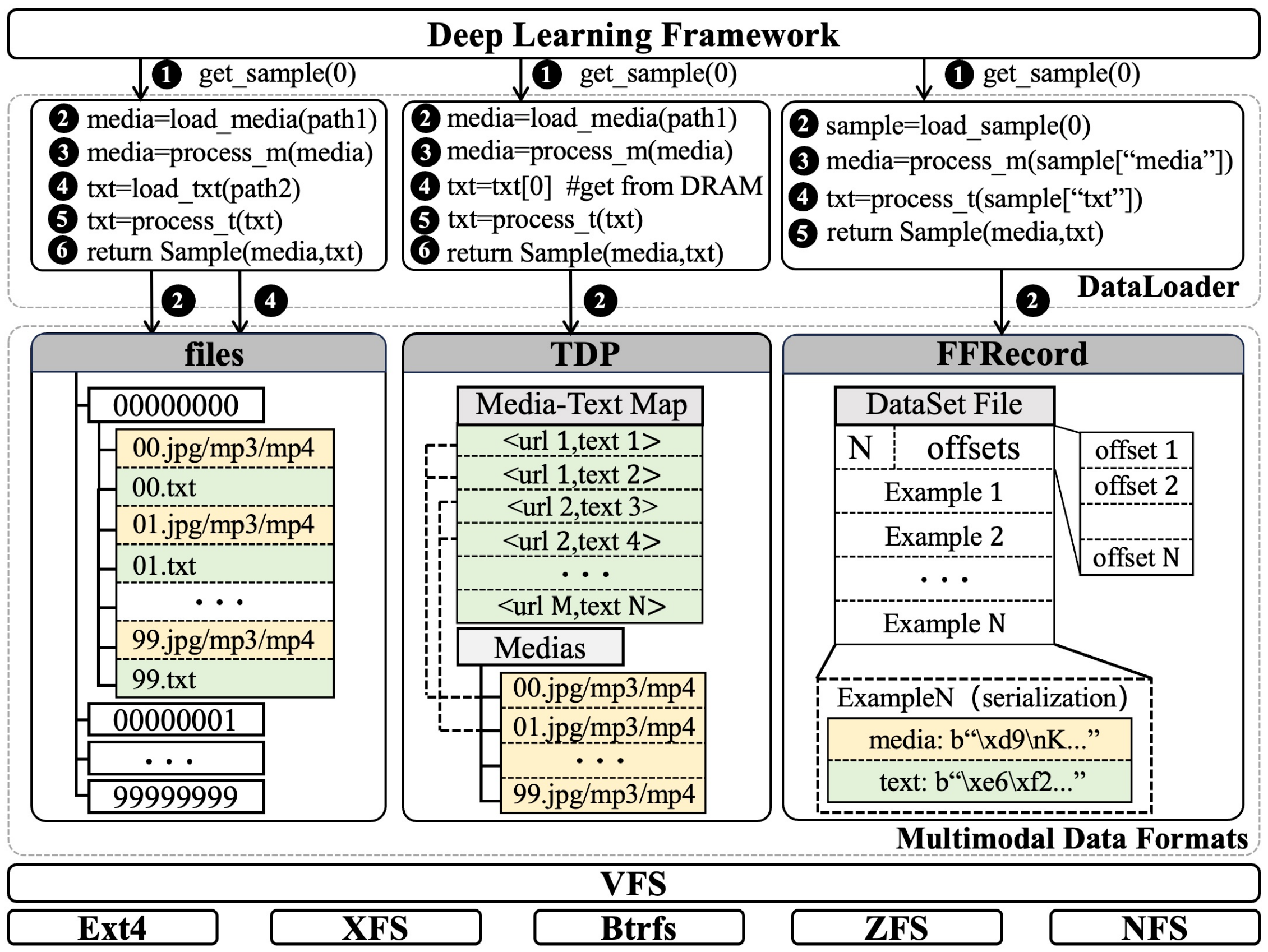}}
\caption{ Data pipelines of different multi-modal storage formats.}
\label{multi_modal storage}
\end{figure}

Figure~\ref{multi_modal storage} illustrates the data pipelines of three commonly used multi-modal storage formats. 

\paragraph{Files}
The \textit{Files} format stores each sample's media and text as separate files organized according to directory hierarchy and naming conventions. A dataset is therefore represented as a very large collection of heterogeneous small files. Sample indices are computed on the fly during training, and each sample is obtained by locating and loading both the media file and the corresponding text file.

\paragraph{Table Data Path (TDP)}
\textit{TDP} aggregates all text entries and alignment metadata into a single large table file (for example CSV, TSV, JSON, or JSONL), while media are stored as individual files. During initialization, the tabular data and the media paths are loaded into memory for indexing, and at training time only the media files are fetched from storage. This design reduces the number of small files compared to \textit{Files}, and replaces media duplication with lightweight path references.

\paragraph{FFRecord}
\textit{FFRecord} serializes each media--text pair into an \texttt{Example} object and writes all objects into a single large file. An offsets array is maintained in the file header to support random access by sample index. During training, the offsets are loaded into memory, and each requested sample is obtained by locating the corresponding \texttt{Example} object.

Although these formats differ in how they organize files and implement indexing, they share a common assumption: in all three designs, the \emph{storage and access unit is the media--text pair}. In other words, the dataset is flattened into a set of independent pairs, and the one-to-many Media2Texts structure is not explicitly represented.

This pair-level granularity is convenient from an interface perspective, since it aligns with the notion of a “sample” in many training pipelines. However, it introduces a structural mismatch with the actual organization of multi-modal data, and as we show next, this mismatch is closely related to four key bottlenecks observed in practice.

\begin{table}[t]
\centering
\caption{Comparison of different data formats. \footnotesize{No existing data format addresses all four issues, as their designs fail to account for the structural characteristics of multi-modal data.}}
\label{tab:dataformat}
\resizebox{\columnwidth}{!}{%
\begin{tabular}{l!{\vrule}cccc}
\toprule
\textbf{Data} & \textbf{Storage} & \textbf{Massive} & \textbf{Cache-} & \textbf{Expensive} \\
\textbf{Format} & \textbf{Redundancy} & \textbf{Small Files} & \textbf{Unfriendly} & \textbf{Data Indexing} \\
\midrule
Files         & \xmark & \xmark & \xmark & \cmark \\
TDP           & \cmark & \xmark & \cmark & \xmark \\
FFRecord      & \xmark & \cmark & \xmark & \cmark \\
\bottomrule
\end{tabular}%
}
\end{table}

\subsection{From Structural Mismatch to System-Level Inefficiencies}
\label{sec3:consequences}

The pair-level granularity faces 4 types of system-level inefficiencies in one-to-many structure, as summarized in Table~\ref{tab:dataformat}.

\subsubsection{Storage Redundancy}
\label{sec3.1}

In a one-to-many multi-modal dataset with a media-to-text ratio of $1{:}N$, a pair-level representation stores each media object $N$ times, pairing it separately with each associated text. Since media objects are much larger than text, the resulting storage consumption grows approximately linearly with $N$, and quickly dominates the overall dataset size.

\begin{figure}[t]
\centerline{\includegraphics[width=1\linewidth]{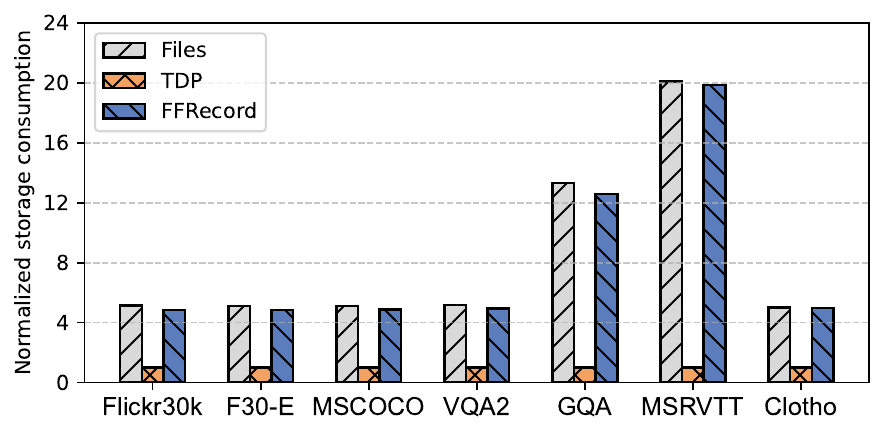}}
\caption{Storage consumption of data formats across datasets (Normalized to TDP). \footnotesize{Both \textit{Files} and \textit{FFRecord} suffer from severe storage redundancy, while \textit{TDP} addresses this issue at the cost of reduced random-sample access performance and increased indexing memory overhead.}}
\label{fig:storage-overhead}
\end{figure}

To quantify this effect, we convert seven representative datasets into \textit{Files}, \textit{TDP}, and \textit{FFRecord} and compare their storage consumption. As shown in Figure~\ref{fig:storage-overhead}, the storage usage of \textit{Files} and \textit{FFRecord} is significantly higher than that of \textit{TDP} across all datasets. This difference is consistent with their design: both \textit{Files} and \textit{FFRecord} manage data at the media--text pair granularity and therefore duplicate media across pairs, while \textit{TDP} separates modalities and replaces duplicated media with lightweight links.

\subsubsection{Cache-Unfriendly Data Layouts}
\label{sec3.2}

Caching is a standard technique to reduce the latency of repeated data accesses. In uni-modal training settings with purely random sampling, the benefit of caching is often limited, because each sample is typically accessed at most once per epoch. In contrast, in multi-modal datasets with a one-to-many structure, the same media object may be accessed multiple times within one epoch through different associated texts, which creates a natural opportunity for cache reuse.

However, effective reuse requires that the cache can recognize and reuse the same media object across different accesses. Pair-level formats such as \textit{Files} and \textit{FFRecord} store redundant copies of the same media either in separate files or in separate serialized objects. Consequently, from the perspective of the underlying cache system, these copies appear as distinct data items, and the cache cannot exploit the latent sharing. In contrast, \textit{TDP} separates modalities and uses references to media files, which allows shared media to be cached and reused across samples.

\begin{figure}[t]
\centerline{\includegraphics[width=1\linewidth]{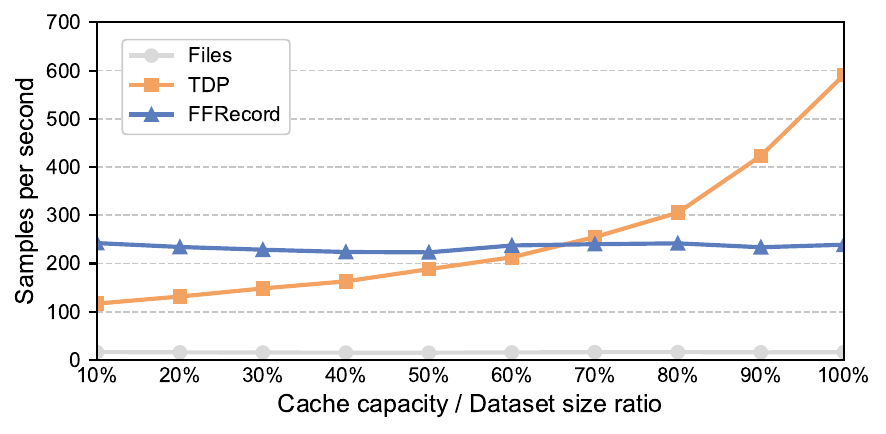}}
\caption{Throughput of each data format on Flickr30k under varying cache-to-dataset size ratios. \footnotesize{Both \textit{Files} and \textit{FFRecord} are cache-unfriendly data formats that cannot exploit the caching system to accelerate access, whereas \textit{TDP} addresses this limitation at the cost of degrading the data format’s inherent random-sample access performance and increasing indexing memory overhead.}}
\label{cache_size}
\end{figure}

To evaluate the impact of cache behavior on sample-level random access performance, we compare \textit{Files}, \textit{TDP}, and \textit{FFRecord} under different cache capacity configurations on Flickr30k. As shown in Figure~\ref{cache_size}, the throughput of \textit{TDP} improves steadily as cache capacity increases, and approaches that of \textit{FFRecord} when the cache size reaches about 70\% of the dataset size. In contrast, the performance of \textit{Files} and \textit{FFRecord} remains almost unchanged as cache capacity increases, which indicates that the default caching mechanism is unable to leverage the potential media reuse within an epoch.

\subsubsection{Indexing Overhead}
\label{sec3.3}

Efficient sample-level random access requires an index that maps logical sample identifiers to their physical storage locations. In current designs, this index is typically constructed at the media--text pair level and is preloaded into memory prior to training in order to support constant-time lookup.

For large-scale multi-modal datasets with a one-to-many structure, a pair-level index necessarily scales with the number of pairs rather than the number of media objects. This effect is further amplified when index structures also embed additional information, such as full-text content and alignment metadata.

% In practice, 
We observe that existing formats adopt a relatively uniform indexing strategy across datasets with diverse characteristics. For example, \textit{TDP} loads all text entries and alignment information into memory, while \textit{FFRecord} maintains a sample-level offsets array. While these designs provide fast lookup, they incur substantial indexing memory overhead that can become a limiting factor when scaling to very large datasets.

\begin{figure}[t]
\centerline{\includegraphics[width=1\linewidth]{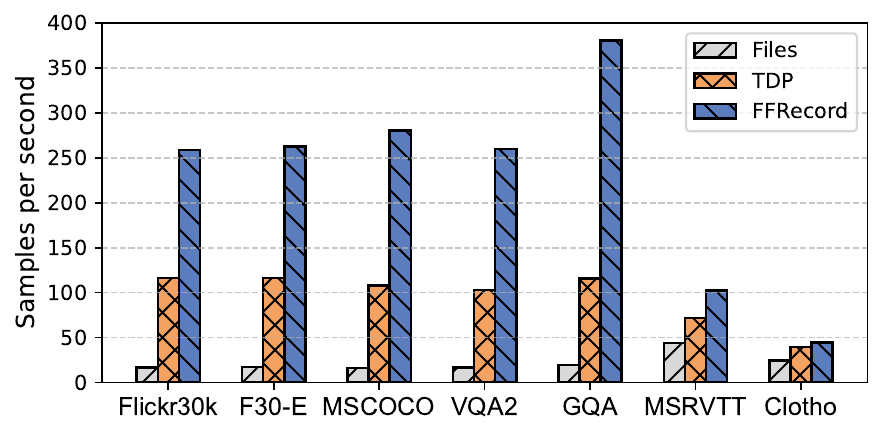}}
\caption{Sample-level random access performance of data formats across datasets with zero cache capacity. \footnotesize{Both \textit{Files} and \textit{TDP} exhibit poor sample-level random access performance due to the large number of small files.}}
\label{Sample-level random access performance}
\end{figure}

\subsubsection{Large Numbers of Small Files}
\label{sec3.4}

The issue of large numbers of small files arises naturally in formats that store each modality of each sample as a separate file. 
For instance, \textit{Files} stores a dataset with $S$ samples and $m$ modalities with $mS$ files.
At web scale, this quickly becomes problematic. For example, WebLI contains over 100 billion image--text pairs; a direct \textit{Files}-style representation would produce more than 200 billion files, which imposes heavy pressure on the file system and on metadata management.

\textit{TDP} partially mitigates this problem by aggregating all text into a single table file and leaving only media as separate files. Assuming a media-to-text ratio of $1{:}N$, this reduces the number of files to approximately $\frac{1}{2N}$ of that required by \textit{Files}.

\textit{FFRecord} serializes all media--text pairs into a single large file and therefore avoids the small-file problem, at the cost of duplicating media and retaining a pair-level layout.

To quantify the effect of file granularity on random sample access, we compare the sample-level throughput of \textit{Files}, \textit{TDP}, and \textit{FFRecord} across seven datasets. As shown in Figure~\ref{Sample-level random access performance}, \textit{FFRecord} achieves substantially higher throughput than both \textit{Files} and \textit{TDP}, which confirms that large numbers of small files can severely degrade random-access performance.

\section{Design of FFSlim}
\label{sec4}

In this section, we describe the design of \textit{FFSlim}, a high-performance and lightweight multi-modal data format that is explicitly built to align with the inherent \textit{one-to-many} structure of modern multi-modal datasets. As shown in Section~\ref{sec3}, existing data formats all operate under a pair-level abstraction that fundamentally conflicts with this structure, which inevitably leads to storage redundancy, cache-unfriendly data layouts, excessive indexing memory usage, and the scalability challenges introduced by massive numbers of small files. FFSlim is designed from the ground up to address these issues through a holistic redesign of data organization, retrieval, and indexing.

The FFSlim system is composed of three tightly coupled components:  
(1) the \textit{Unified Lightweight High-Performance File Format} (ULHFF), an on-disk representation that physically encodes multi-modal data at the \textit{Media2Texts} granularity;  
(2) the \textit{Adaptive Sample Retrieval} (ASR) layer, which provides pair-level random access atop Media2Texts objects; and  
(3) the dataset \textit{Conversion Tool}, which transforms existing formats into ULHFF and constructs the corresponding index structures.  
We now expand each component in detail.

\subsection{ULHFF: A Unified Lightweight High-Performance File Format} 
\label{sec4.2}
\begin{figure}
\centerline{\includegraphics[width=1\linewidth]{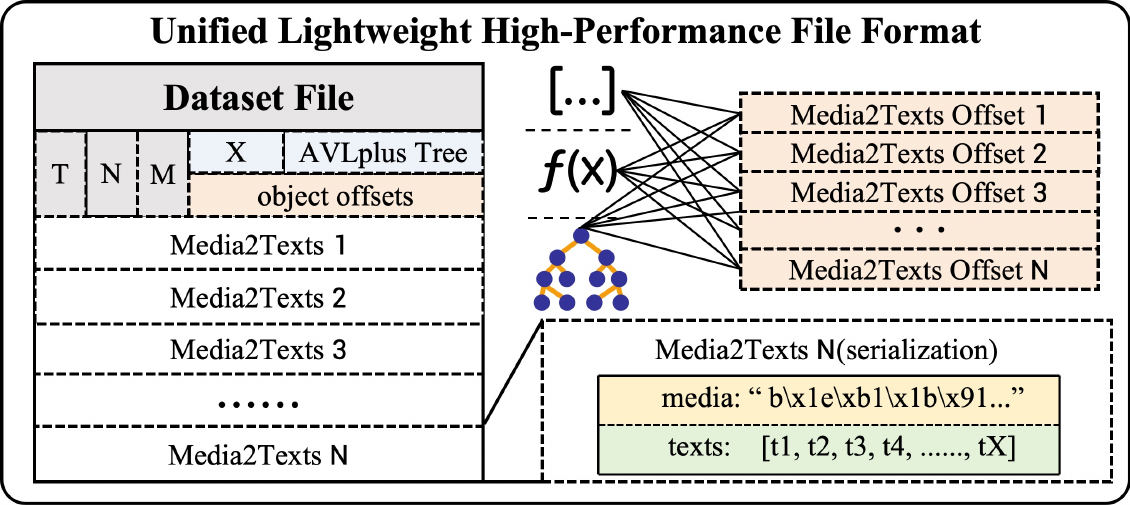}}
\caption{Overview of a unified, lightweight, and high-performance file format (ULHFF).}
\label{Overview of ULHFF}
\end{figure}

To eliminate storage redundancy, improve cache behavior, and avoid the large-number-of-small-files problem, we design the \textit{Unified Lightweight High-Performance File Format} (ULHFF). ULHFF adopts the \textit{Media2Texts} granularity as its basic storage unit and organizes all data into a single, contiguous file. This design preserves the inherent one-to-many structure of multi-modal datasets and provides a stable physical foundation for efficient random access. It also matches the empirical observation that, in most multi-modal corpora, the media object is far larger than any of its associated text entries; exploiting this asymmetry is key to achieving both storage and caching efficiency.

ULHFF encodes each media object together with all its associated text entries in a single \textit{Media2Texts} object. Storing shared media exactly once removes cross-modal redundancy, while co-locating media with its texts restores spatial locality and allows the OS page cache to exploit repeated access within an epoch. In typical one-to-many settings, this co-location significantly amplifies cache reuse. 
A limitation of this approach is that if the media object is not substantially larger than the text entries, the cost of loading the entire Media2Texts object may outweigh the benefits of locality. In practice, however, most multi-modal datasets exhibit a strong media-dominant size ratio, making ULHFF well-suited for real workloads.

As the figure~\ref{Overview of ULHFF} shows, the ULHFF file consists of a header and a body. The header stores essential metadata and the index structure required for random access; the body contains a sequence of $M$ serialized Media2Texts objects. Each object stores its media under \texttt{"media"} and its text entries under \texttt{"texts"}, forming a compact representation that is independent of the original dataset format. This simple layout not only avoids the filesystem overhead of managing a massive number of small files but also improves I/O predictability by exposing a single, contiguous byte stream to the storage subsystem.

The ULHFF header contains six logical fields, of which only a subset is activated depending on the dataset type.  
\textit{T} records the pairing paradigm (\textit{One-to-One}, \textit{Fixed One-to-Many}, \textit{Hybrid One-to-Many}, or uni-modal). 
\textit{N} and \textit{M} specify the number of logical media--text pairs and the number of Media2Texts objects.  
An \textit{object offsets} array of length $M$ records the starting byte offset of each Media2Texts object in the body, enabling direct offset-based retrieval.  
For fixed-ratio datasets, ULHFF stores \textit{X}, the fixed media-to-text ratio, which supports constant-time arithmetic indexing.  
For hybrid datasets with non-uniform media-to-text ratios, ULHFF embeds an \textit{AVLplus Tree}---a compact, balanced-tree index. Each node stores a \texttt{(start\_idx, end\_idx, offset)} triplet, where \texttt{start\_idx} and \texttt{end\_idx} bound the sample IDs encapsulated in the corresponding \texttt{Media2Texts} object, and \texttt{offset} points to its physical byte location. 
AVLplus enables $O(\log M)$ lookup while keeping index memory proportional to the number of Media2Texts objects rather than the number of media–text pairs. This trades a modest increase in lookup complexity for a substantial reduction in index memory footprint, which is particularly beneficial for large-scale hybrid datasets.

The file body contains exactly $M$ serialized Media2Texts objects arranged consecutively. This single-file organization simplifies I/O paths, ensures cache-friendly object placement, and eliminates the scalability issues caused by directory traversal and per-file metadata management. ULHFF therefore provides not only a more faithful representation of multi-modal data, but also a more robust and performant foundation for storage and retrieval compared to pair-level formats such as Files, TDP, and FFRecord.

\subsection{ASR: Adaptive Sample Retrieval}
\label{sec4.4}
\begin{figure}
\centerline{\includegraphics[width=\linewidth]{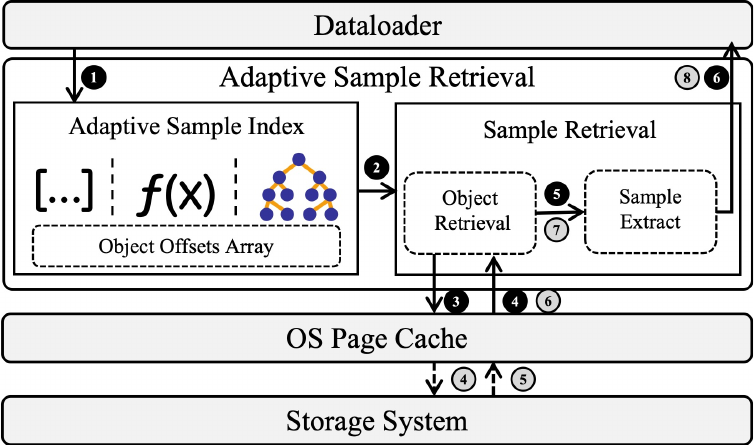}}
\caption{Workflow of adaptive sample retrieval.}
\label{workflow_adaptive}
\end{figure}

While ULHFF organizes data at the \textit{Media2Texts} granularity, training pipelines continue to operate at the media–text–pair level. The goal of \textit{Adaptive Sample Retrieval} (ASR) is therefore to bridge this granularity gap and provide efficient, fully random pair-level access without negating the structural advantages of the Media2Texts layout. ASR is designed with the observation that media objects in typical multi-modal datasets are far larger than their associated text entries; leveraging this asymmetry is essential to achieving high-throughput random access while maintaining a small indexing footprint.

ASR is built on two principles.  
First, FFSlim retrieves complete \textit{Media2Texts} objects rather than individual media–text pairs. Because media objects dominate object size, the additional cost of reading extra text entries is negligible. Once loaded, the object naturally benefits from OS page caching, enabling subsequent accesses to any of its text entries to be served from memory. This design is particularly effective in one-to-many scenarios, where different text entries referencing the same media are frequently queried within an epoch. A limitation of this strategy is that when the media–text size ratio is small, the read amplification introduced by fetching the entire object becomes more noticeable; in such cases, caching may not fully offset the additional cost.

Second, ASR adaptively selects an indexing structure based on the dataset’s pairing paradigm. This avoids a one-size-fits-all reliance on per-pair hash tables, significantly reducing memory usage while preserving lookup efficiency. In particular, for hybrid one-to-many datasets, ASR employs the \textit{AVLplus} index, which reduces indexing memory by 63.64--85.82\% compared to hash-based designs while retaining log\-arithmic-time lookup. This trade-off is deliberate: for large hybrid datasets, the memory savings from indexing at Media2Texts granularity far outweigh the modest increase in lookup complexity.

Figure~\ref{workflow_adaptive} shows the retrieval pipeline. When the \textit{DataLoader} requests sample $i$ (\ding{172}), ASR first consults the in-memory index to locate the Media2Texts object containing this sample and its intra-object offset (\ding{173}). The object is then retrieved from the ULHFF body via an offset-based read (\ding{174}). If the object is already in the OS page cache, the read is served immediately (\ding{175}); otherwise, it is fetched from storage (\ding{175}--\ding{177}). The requested media–text pair is finally extracted from the loaded Media2Texts object and returned to the training pipeline (\ding{176}--\ding{179}).

During initialization, ASR constructs the \textit{Adaptive Sample Index} by parsing the ULHFF header. For one-to-one or uni-modal datasets, the logical sample ID is directly used as the index into the \textit{object offsets} array, and the intra-object offset is always zero. Fixed one-to-many datasets exploit the fixed media-to-text ratio $X$: simple arithmetic decomposes a logical sample ID into the corresponding Media2Texts index and text offset, after which the byte address is obtained from the \textit{object offsets} array. Hybrid one-to-many datasets require a more flexible structure, and ASR uses the \textit{AVLplus} tree embedded in the header. Each AVLplus node stores the logical sample range covered by a Media2Texts object, enabling ASR to locate the correct object and compute the intra-object offset via a balanced-tree search.

The AVLplus lookup procedure is as follows. Given a logical sample ID, ASR compares it with the root node’s \texttt{start\_idx} and \texttt{end\_idx} fields. If the ID lies within the range, ASR returns the node’s Media2Texts address and computes the intra-object offset as \texttt{id - start\_idx}. If the ID is smaller, ASR descends into the left subtree; otherwise, it descends into the right subtree. Because AVLplus is height-balanced, lookup latency is $O(\log M)$, where $M$ is the number of Media2Texts objects. Since the tree is memory-resident and $M$ is typically much smaller than the total number of media–text pairs, this cost is negligible in practice.

Together, object-level retrieval and adaptive indexing allow ASR to provide efficient, pair-level random access without undermining the Media2Texts-centric organization introduced by ULHFF. The design explicitly trades a slightly more complex lookup for dramatically reduced indexing memory footprint, enabling FFSlim to scale gracefully to large one-to-many and hybrid datasets while retaining throughput comparable to or better than pair-level formats. This decoupling of logical sample identity from physical storage granularity is fundamental to FFSlim’s performance and system efficiency.

\subsection{Conversion Tool}
\label{sec4.3}

\begin{figure}[t]
\centerline{\includegraphics[width=\linewidth]{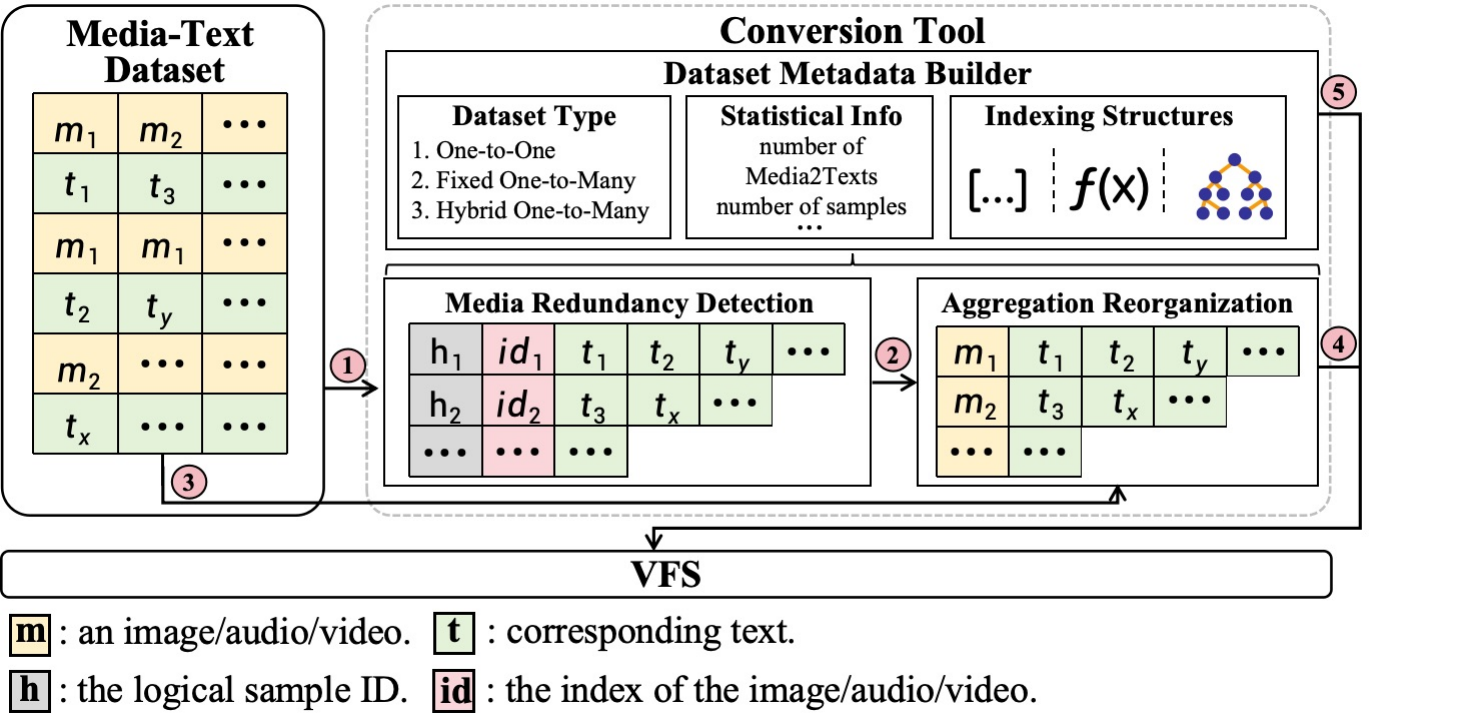}}
\caption{Workflow of the conversion tool.}
\label{workflow_conversion_tool}
\end{figure}

To make FFSlim directly usable on existing datasets, we provide a \textit{conversion tool} that transforms legacy formats such as Files, TDP, and FFRecord into the ULHFF structure. The conversion process is performed once per dataset and produces a fully self-contained ULHFF file with deduplicated Media2Texts objects and dataset-specific indexing metadata. As shown in Figure~\ref{workflow_conversion_tool}, the tool operates in three coordinated stages: detecting media redundancy, aggregating pairs into Media2Texts objects, and constructing the final dataset metadata and index structures.

The conversion begins with \textit{Media Redundancy Detection}. It iterates over all media–text pairs in the original dataset (\ding{172}), computes a content hash for each media object (\eg, \textit{$h_1$}), and maintains a hash table that maps each hash value to a representative media index and its associated text entries. 
% Only media indices are stored; 
Raw media bytes are not retained in memory. This design keeps the memory footprint proportional to the number of distinct media objects rather than the number of pairs. When multiple pairs reference the same media content, their text entries are accumulated under the same hash key.

Once redundancy detection identifies a complete set of media clusters (\ding{173}), the \textit{Aggregation Reorganization} module processes each cluster. It retrieves the corresponding media object from storage using its representative index (\ding{174}), aggregates it with all collected text entries, and forms a \textit{Media2Texts} object. The object is then serialized directly into the ULHFF body in the order processed, and its starting byte offset is recorded (\ding{175}). This streaming-friendly procedure allows the tool to scale to large datasets without materializing all Media2Texts objects in memory at once.

After all Media2Texts objects have been serialized, the \textit{Dataset Metadata Builder} finalizes the ULHFF header (\ding{176}). It examines the number of text entries associated with each Media2Texts object to determine the dataset type.
% —\textit{One-to-One}, \textit{Fixed One-to-Many}, or \textit{Hybrid One-to-Many}. 
Based on the type, it constructs the corresponding dataset-specific index structure: trivial array indexing for one-to-one datasets, fixed-ratio arithmetic indexing for fixed one-to-many datasets, or the AVLplus tree for hybrid datasets. The builder also records dataset statistics, including the logical pair count $N$, the number of Media2Texts objects $M$, and the full \textit{object offsets} array. 

The output of the conversion tool is a single ULHFF file containing a compact, deduplicated, and index-equipped representation of the original dataset. This file becomes the sole input to ASR during training, ensuring that all subsequent random sample accesses benefit from Media2Texts-level organization and lightweight indexing.

\begin{figure}
\centerline{\includegraphics[width=1\linewidth]{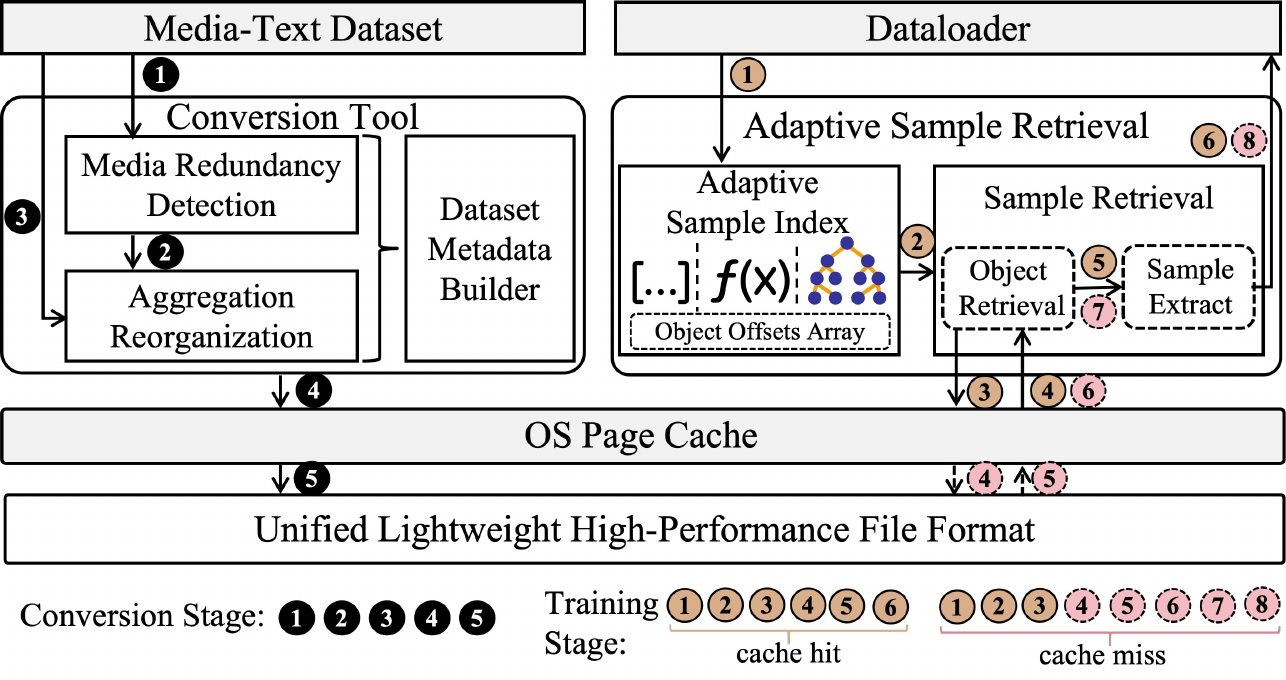}}
\caption{Overview of FFSlim.}
\label{Overview of FFSlim}
\end{figure}

\subsection{Implementation and Training Integration}
\label{sec4.5}

FFSlim is a practical storage and retrieval subsystem that integrates seamlessly with existing deep learning workflows. 

\textbf{Workflow.}
FFSlim operates in two stages: conversion and training. 
During conversion, the system traverses the dataset in its original format, detects duplicate media objects, aggregates their associated text entries, constructs \textit{Media2Texts} objects, and serializes them into the ULHFF file body. The dataset type is then inferred, and the corresponding lightweight index structure is generated and written into the file header.  
During training, the framework issues pair-level sample requests through the standard \texttt{DataLoader} interface. ASR resolves each logical sample ID to a Media2Texts object and intra-object offset using the in-memory index, retrieves the object via a single offset-based read (benefiting from OS page caching whenever possible), and extracts the requested media–text pair before returning it to the training pipeline. The overall process is depicted in Figure~\ref{Overview of FFSlim}.

\textbf{Support for diverse training strategies.}
Multi-modal training pipelines use different data-access strategies depending on model architecture and task. FFSlim naturally supports them because ASR always retrieves full Media2Texts objects.  
When a training regime samples all text entries associated with each media within an epoch, the initial lookup loads the Media2Texts object and subsequent accesses exploit cache locality. When training samples a single text per media per epoch, ASR simply extracts the required entry and discards the rest without additional cost. For settings that concatenate multiple text entries into a single long sequence paired with the same media, ASR loads the Media2Texts object and allows the upper layer to construct the desired sequence directly. These behaviors require no modification to the ULHFF format or indexing logic.

\textbf{Prototype.}
We implement FFSlim in PyTorch~2.7.0 by providing a \texttt{FFSlimDataset} class that conforms to \texttt{torch.utils\-.data.Dataset}. The class parses ULHFF files, constructs the adaptive sample index during initialization, and serves sample requests through ASR. We use \texttt{pickle} for serializing metadata and Media2Texts objects, \texttt{hashlib} for media hashing, and standard \texttt{os} interfaces for file operations. The conversion tool is launched via a simple shell script \texttt{conversion.sh} that accepts the input dataset path and outputs the corresponding ULHFF file.

\section{Evaluation}
\label{sec5}

We evaluate FFSlim across a comprehensive set of metrics, including data loading performance, write throughput, storage consumption, index memory footprint, and sensitivity to cache capacity. Our experiments span seven representative multi-modal datasets and compare FFSlim with three widely used baseline formats. We further examine the tradeoff between indexing efficiency and memory overhead, present an end-to-end case study on real training workloads, and conduct ablation studies to isolate the contributions of ULHFF and ASR.

\subsection{Experimental Setup}
\label{sec:exp_setup}

\textbf{Metrics.}
We use four metrics to assess storage and retrieval performance.  
Data loading performance is measured as the number of samples loaded per second under fully randomized access, averaged across one full training epoch.  
Write performance is defined as the throughput of writing the entire dataset to storage; we report the average throughput when copying the dataset to a parallel file system.  
Storage consumption is measured as the sum of on-disk data size and estimated filesystem metadata overhead, where metadata usage is approximated as $4$KB per file multiplied by the number of files.  
Index memory footprint refers to the runtime memory consumed by in-memory indexing structures during training.

\textbf{Workloads.}
We benchmark all formats across seven representative multi-modal datasets spanning image–text, video\-–text, and audio–text modalities (Table~\ref{tab:dataset}). For each dataset, we evaluate performance under fully random sample loading. The effective OS page-cache capacity available to the training job is set to $x\%$ of the dataset size by default. Because cache-to-dataset ratios vary widely in real deployments, we sweep $x$ from 10\% to 100\% across ten configurations and report the mean loading performance. Write performance is measured independently by transferring each dataset from local storage to a parallel file system.

\textbf{Baselines.}
We evaluate FFSlim against three widely used data formats: raw files~\cite{radford2021learning}, TDP~\cite{alayrac2022flamingo}, and FFRecord~\cite{ffrecord}. We deliberately exclude WebDataset~\cite{webdataset}, TFRecord~\cite{tfrecord}, and MDS~\cite{mosaicml2022streaming} from our baselines. To maximize loading throughput, these formats compromise model accuracy by utilizing pseudo-random sequential streaming, which structurally precludes the user-defined, fully random access workloads required in our evaluation.

% 实验平台
\textbf{Platform.} 
All experiments are conudcted on a training server equipped with $2\times$ AMD EPYC 7742 CPUs (64cores), 512~GB DRAM, 10~Gbps Ethernet, 8$\times$~ H200 GPUs, and 2.9~TB NVMe SSD formatted with ext4. The system runs Ubuntu~24.04 with Linux kernel~v6.8.0. All multi-modal models are trained with PyTorch~2.7.0, with datasets stored on a 3FS~\cite{3fs2024} parallel file system located in the same data center.

\subsection{Performance Results}
\begin{figure}[t]
\centerline{\includegraphics[width=\linewidth]{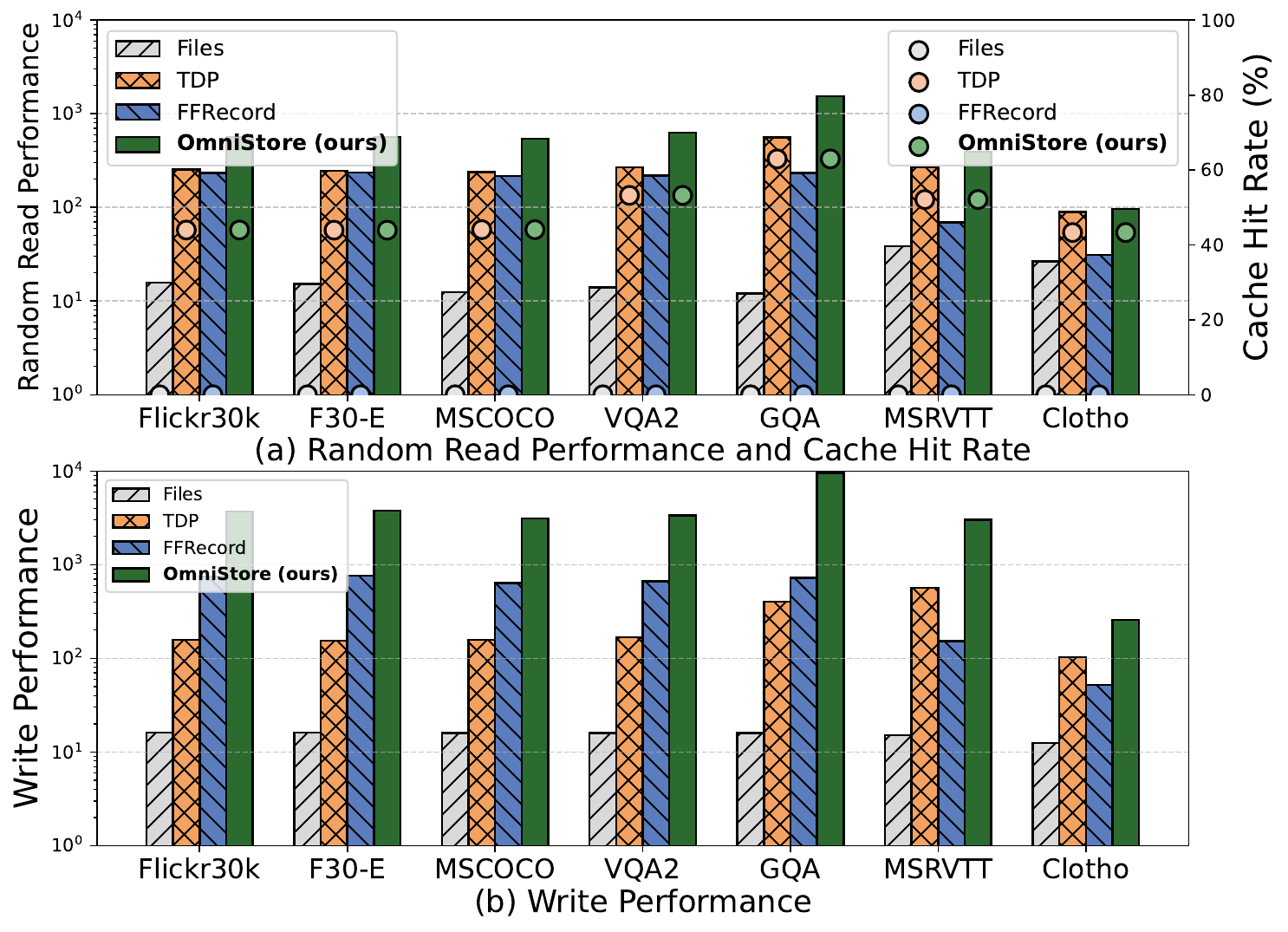}}
\caption{Data loading performance(samples/second) and dataset write throughput(samples/second) across diverse multi-modal datasets.}

\label{figure9}
\end{figure}

In this section, we evaluate FFSlim’s end-to-end performance across representative multi-modal datasets. We focus on two key dimensions that dominate real-world training efficiency: data loading throughput and write throughput. Together, these metrics capture both the runtime behavior of sample retrieval during training and the cost of constructing datasets at scale.
\label{sec6.2}
% 数据加载性能
\subsubsection{Data loading performance}
% 实验结果与分析
% 总体结果 --> 与baseline量化对比(范围/平均) --> 分析原因 -->  不同benchmark的差异性分析 --> 其他baseline差的原因(非重点)
As shown in Figure~\ref{figure9}, \textit{FFSlim} consistently achieves the highest data-loading throughput, delivering an average speedup of $2.07\times$ over the best-performing baseline.
These gains stem from the co-design of its sample organization and retrieval strategies. First, \textit{FFSlim} encodes one-to-many pairing paradigms as \textit{Media2Texts} objects and manages samples at this granularity, grouping all texts associated with the same media. This layout is inherently cache-friendly and accelerates repeated media accesses. Second, instead of fetching individual media-text pairs, \textit{FFSlim} retrieves an entire \textit{Media2Texts} object and extracts the requested pair, which further improves access locality and speeds up subsequent accesses to other media-text pairs already resident in the page cache. Finally, by storing the entire dataset in a single unified file, \textit{FFSlim} eliminates the overhead of managing large numbers of small files, providing an additional boost to data-loading performance.

% 不同数据集特征对Baseline的影响(以FFSlim为主)
Compared with \textit{FFRecord}, \textit{FFSlim} achieves larger gains on \textit{GQA} and \textit{MSRVTT}, as their higher media-to-text ratios increase the likelihood that cached \textit{Media2Texts} objects are reused within an epoch, thereby boosting cache hit rates and data-loading throughput. In contrast, \textit{FFRecord} employs a cache-unfriendly layout and thus cannot exploit page-cache locality to accelerate loading.
Compared with \textit{TDP}, \textit{FFSlim} shows smaller gains on \textit{MSRVTT} and \textit{Clotho}, since these datasets contain relatively few but large media objects, making the small-file problem in \textit{TDP} less severe and leaving limited headroom for improvement. Finally, \textit{Files} performs worst across all datasets, as it suffers the most from the issue of large number of small files , is cache-unfriendly, and further incurs two I/Os per sample.

% 写入性能
\subsubsection{Write performance}
% 解释为什么要测试写入性能
Although the training process only involves read operations, datasets must first be written to the storage system, making write performance an essential consideration for data formats. 

% 实验结果与分析
As shown in Figure~\ref{figure9}, \textit{FFSlim} delivers the highest write throughput across seven representative multi-modal datasets, delivering an average speedup of $8.26\times$ over the best-perform\-ing baseline. 
Although datasets are written only once prior to training, severe write slowdowns on large-scale datasets can still become a critical bottleneck that delays the entire training pipeline. The performance gain stems from \textit{FFSlim}'s unified and lightweight file format: multiple text entries share a single media object within a \textit{Media2Texts} instance, substantially reducing the total volume of data written. Moreover, by organizing the entire dataset into a single file, \textit{FFSlim} converts massive small-file random writes into a sequential write to a single file, further improving write efficiency.

% 不同数据集特征对Baseline的影响(以FFSlim为主)
Compared to \textit{FFRecord}, \textit{FFSlim} delivers more substantial improvements on \textit{GQA} and \textit{MSRVTT}. The higher media-to-text ratios in these datasets exacerbate data redundancy in \textit{FFRecord}, resulting in an $N\times$ larger write volume than \textit{FFSlim} (where $N$ denotes the media-to-text ratio). Conversely, compared to \textit{TDP}, \textit{FFSlim} exhibits smaller gains on \textit{MSRVTT} and \textit{Clotho}. Because these datasets contain relatively few but large media objects, \textit{TDP}'s small-file overhead is less pronounced, leaving limited room for optimization.

Notably, \textit{TDP} outperforms \textit{FFRecord} on \textit{MSRVTT} and \textit{Clotho}, but falls behind on the other datasets due to a shift in I/O bottlenecks. On \textit{MSRVTT} and \textit{Clotho}, the larger media sizes prevent severe write degradation. Consequently, total write volume becomes the primary bottleneck, allowing \textit{TDP} to benefit from its redundancy elimination. On other datasets, however, the massive number of small files drastically degrades \textit{TDP}'s write bandwidth. In these cases, write bandwidth---rather than write volume---becomes the bottleneck, rendering \textit{TDP} slower than \textit{FFRecord} despite writing fewer bytes. Furthermore, \textit{Files} consistently performs the worst across all datasets, as it suffers from both unoptimized write volumes and extensive small-file random writes.

% 存储开销
\subsection{Storage Consumption}
% 实验结果与分析
% 我们方案的具体开销+原因 --> 与baseline量化对比(范围/平均) --> 分析原因 -->  异常补充分析 --> 其他baseline差的原因
The storage consumption of \textit{FFSlim} equals the dataset size without duplicated media and links, and without excessive metadata. This comes from \textit{FFSlim}'s unified and lightweight file format, where multiple text entries share the same media object within a \textit{Media2Texts} instance, significantly reducing storage redundancy, and its efficient organization, which stores the entire dataset as a single file, substantially lowering the metadata overhead caused by large number of small files.

\begin{figure}[t]
\centerline{\includegraphics[width=\linewidth]{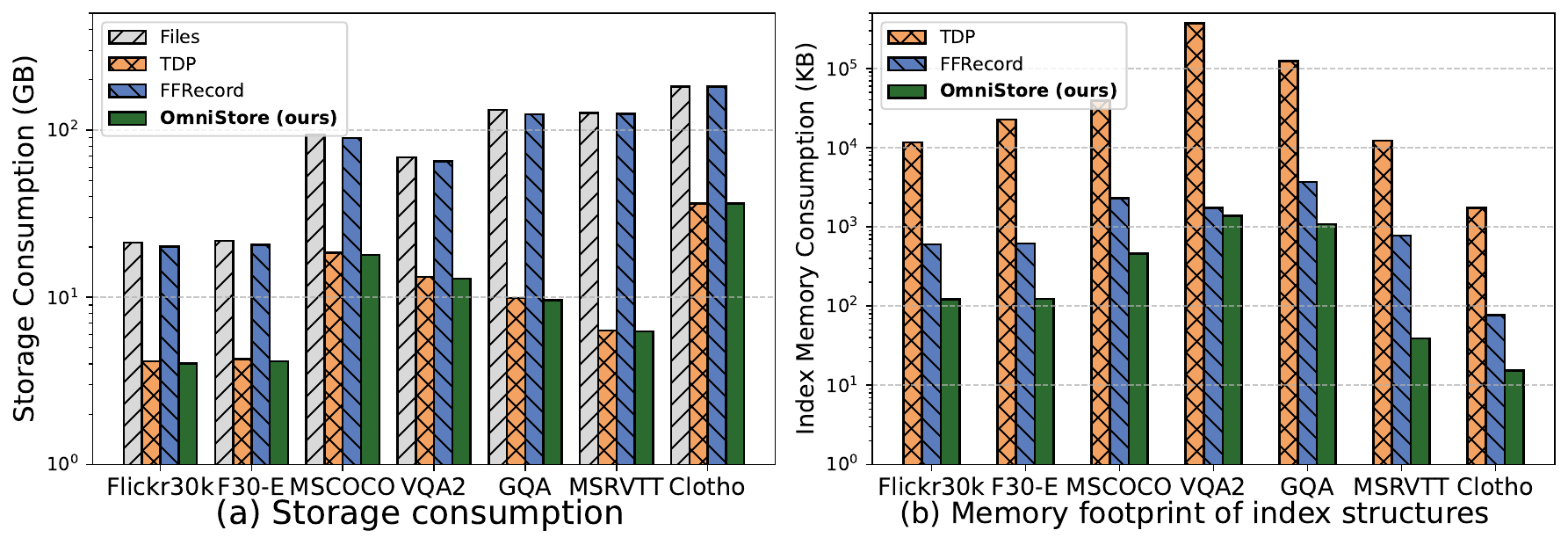}}
\caption{Storage consumption and memory footprint of index structures across diverse multi-modal datasets.}
\label{figure10}
\end{figure}

As shown in Figure~\ref{figure10}, \textit{FFSlim} incurs the smallest storage footprint across all seven datasets, achieving an average reduction of 2.09\% compared to the best-performing baseline.
It is worth noting that although \textit{FFRecord} delivers strong performance, its storage consumption is very high, averaging 8.28$\times$ that of \textit{FFSlim} due to its coarse cross-modal alignment strategy. \textit{TDP} has comparable overhead to \textit{FFSlim} (1.02$\times$) but inferior overall performance. \textit{Files} suffer from both poor performance and high storage consumption, averaging 8.58$\times$ that of \textit{FFSlim}.

% 内存开销
\subsection{Memory Footprint of Index Structures}
\label{sec6.5}
% 为什么索引内存开销重要
Although the index size is small relative to the entire dataset, the dataset itself is much larger than memory, causing the index to occupy a substantial portion of memory and making index overhead an essential factor for data formats. Notably, high memory consumption can impact training by triggering OS-level swapping, which offloads training data and parameters to SSDs and further increases training time.

% 实验结果与分析
% 总体结果 --> 与baseline量化对比(范围/平均) --> 分析原因 -->  异常补充分析 --> 其他baseline差的原因
On average, the index memory overhead of \textit{FFSlim} is only 0.0043\% of the dataset size; even when the dataset is 1000$\times$ larger than memory, the index consumes merely 4.3\% of memory.
This efficiency stems from \textit{FFSlim}'s sample management strategy---organizing samples at the \textit{Media2Texts} granularity---which drastically reduces the number of index nodes, and from its \textit{Adaptive Sample Index Structure}, which selects lightweight and efficient index organizations based on dataset characteristics. For hybrid one-to-many datasets, we further introduce the AVLplus Tree, which trades a negligible amount of sample-loading performance for substantial reductions in index memory footprint.
Moreover, \textit{FFSlim}'s data format naturally avoids embedding textual sample content into index structures, thereby eliminating the high index-memory overhead incurred by \textit{TDP}.

As shown in Figure~\ref{figure10}, \textit{FFSlim} achieves a much lower index memory footprint across all seven datasets, reducing it by 72.23\% and 99.26\% on average compared to \textit{FFRecord} and \textit{TDP}. 
\textit{TDP} has the highest index memory overhead, averaging 168.01$\times$ that of \textit{FFSlim}, because it embeds all text data into the index and builds array indices at the sample level, requiring many objects to be indexed. \textit{FFRecord} also has substantial overhead, averaging 6.37$\times$ that of \textit{FFSlim} due to sample-level array indexing. \textit{Files} incur minimal index memory overhead, as their index is computed dynamically, but suffer from poor I/O performance and high storage consumption.

% Cache size对性能的影响
\subsection{Impact of Cache Capacity}
\begin{figure}
\centerline{\includegraphics[width=\linewidth]{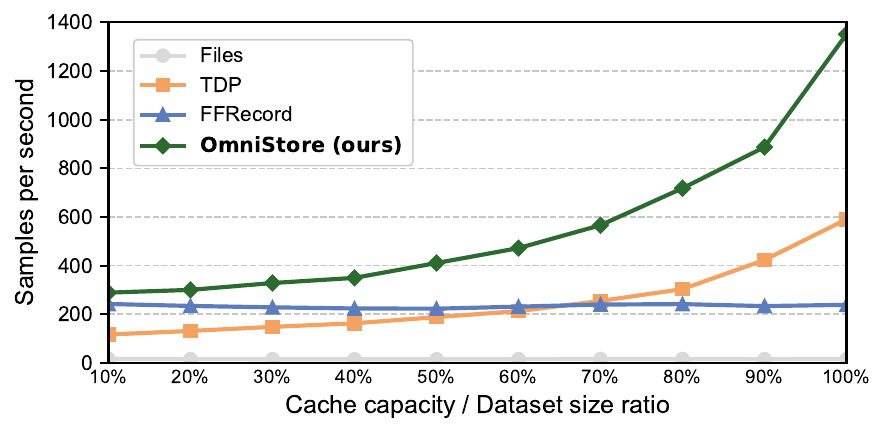}}
\caption{Data loading performance on Flickr30k under different cache capacity.}
\label{cache capacity}
\end{figure}

Figure~\ref{cache capacity} shows the data loading performance of \textit{FFSlim} under different cache capacity configurations on Flickr30k.  
We observe that \textit{FFSlim} consistently outperforms the three baselines across all cache capacity settings. 
Specifically, \textit{FFSlim} achieves an average data loading performance improvement of 2.25$\times$ compared to the best-performing baseline on \textit{Flickr30k}. 
These results demonstrate that \textit{FFSlim} remains effective across a wide range of cache capacity configurations.

\textit{Files} and \textit{FFRecord} have stable data loading performance under different cache capacities, and are thus considered cache-unfriendly.  
\textit{TDP} generally performs poorly, but its performance gradually improves with increased cache capacity, surpassing \textit{FFRecord} when the cache space exceeds 70\% of the dataset size, making it a cache-friendly format.

\subsection{Tradeoff Between Index Efficiency and Memory Overhead}
Table~\ref{tab:tradeoff} reports the data loading performance and index memory overhead of \textit{FFSlim} under different index structures across multiple datasets. We observe that the adaptive index reduces memory usage by 87.24\% on average while incurring only a 0.26\% average performance loss compared to the hash index. This improvement stems from selecting or designing index structures according to dataset characteristics to achieve both high performance and lightweight indexing. For one-to-one and fixed one-to-many datasets, it adopts a hash function combined with a \textit{Media2Texts} offset array, which significantly lowers index memory overhead without sacrificing performance. For hybrid one-to-many datasets (VQA2, GQA), we introduce the \textit{AVLplus Tree} with a \textit{Media2Texts} offset array, trading marginal performance drops (0.51\%, 1.35\%) for substantial reductions in index memory usage (63.64\%, 85.82\%).

\begin{table}[t]
\centering
\caption{Index performance and memory overhead across datasets. \footnotesize{The adaptive index trades negligible performance overhead for substantial reductions in index memory usage.}}
\label{tab:tradeoff}
\resizebox{\columnwidth}{!}{%
\begin{tabular}{l|c|c|c|c} 
\toprule
\multirow{2}{*}{\raisebox{-0.5ex}{Workload}} & \multicolumn{2}{c|}{Samples per Second} & \multicolumn{2}{c}{Memory footprint of index(KB)} \\
\cmidrule(lr){2-3} \cmidrule(lr){4-5}
& Hash index 
& \textbf{Adaptive index (ours)}  
& Hash index 
& \textbf{Adaptive index (ours)} \\
\midrule
Flickr30k & 567.35   & 567.45   & 1,332.63 & 121.14  \\
F30-E     & 565.45   & 565.55   & 1,365.68 & 124.15  \\
MSCOCO    & 541.82   & 541.88   & 5,069.11 & 460.77  \\
VQA2      & 632.69   & 629.46   & 3,813.54 & 1,386.74  \\
GQA       & 1,563.88 & 1,542.79 & 7,638.04 & 1,083.41  \\
MSRVTT    & 394.83   & 394.82 	& 1,601.56 & 39.06  \\
Clotho    & 96.27 	 & 96.28 	& 169.21   & 15.38 \\
\bottomrule
\end{tabular}
}
\end{table}

\subsection{End-to-End Case Study}
% 实验目的与实验方法
To evaluate \textit{FFSlim} in an end-to-end training setting, we adopt LAVIS~\cite{lavis}, a one-stop library for language–vision intelligence. Its training workloads are summarized in Table~\ref{tab:end-to-end-workload}. Here, \textit{Parameters} denotes the total number of model parameters, whereas \textit{Trainable Parameters} refers to those actually updated during training. In typical multi-modal training pipelines, the LLM or ViT backbone is frozen, and only the intermediate projection layers are optimized for cross-modal alignment and fusion, resulting in a much smaller set of trainable parameters. When models become sufficiently large, training may hit GPU memory limits; thus, gradient checkpointing is commonly employed to trade additional computation for reduced memory usage. By default, LAVIS organizes multi-modal datasets using the \textit{TDP} format(SOTA). We convert these datasets into the \textit{FFSlim} format and integrate them into LAVIS to conduct end-to-end training.

\begin{table}[t]
    \centering
    \caption{Model and dataset configuration.}
    \label{tab:end-to-end-workload}
    \resizebox{\columnwidth}{!}{%
    \begin{tabular}{lccccc}
    \toprule
    \textbf{Model} & \textbf{Dataset} & \textbf{Parameters} & \makecell[c]{\textbf{Trainable} \\ \textbf{Parameters}} & \makecell[c]{\textbf{Freeze} \\ \textbf{VIT}} & \makecell[c]{\textbf{Gradient} \\ \textbf{Checkpoint}} \\
    \midrule
    BLIP-Base~\cite{blip} & MSCOCO & 224M & 224M & \xmark & \xmark \\
    BLIP-Large~\cite{blip} & MSCOCO & 446M & 446M & \xmark & \xmark \\
    BLIP2-OPT2.7-F~\cite{li2023blip} & MSCOCO & 3.7B & 107M & \cmark & \xmark \\
    BLIP2-OPT6.7-F~\cite{li2023blip} & MSCOCO & 7.8B & 108M & \cmark & \xmark \\
    BLIP2-OPT2.7~\cite{li2023blip} & MSCOCO & 3.7B & 1.1B & \xmark & \xmark \\
    BLIP2-OPT2.7-GC~\cite{li2023blip} & MSCOCO & 3.7B & 1.1B & \xmark & \cmark \\
    BLIP2-OPT6.7-GC~\cite{li2023blip} & MSCOCO & 7.8B & 1.1B & \xmark & \cmark \\
    \bottomrule
    \end{tabular}
    }
\end{table}

\begin{figure}
\centerline{\includegraphics[width=\linewidth]{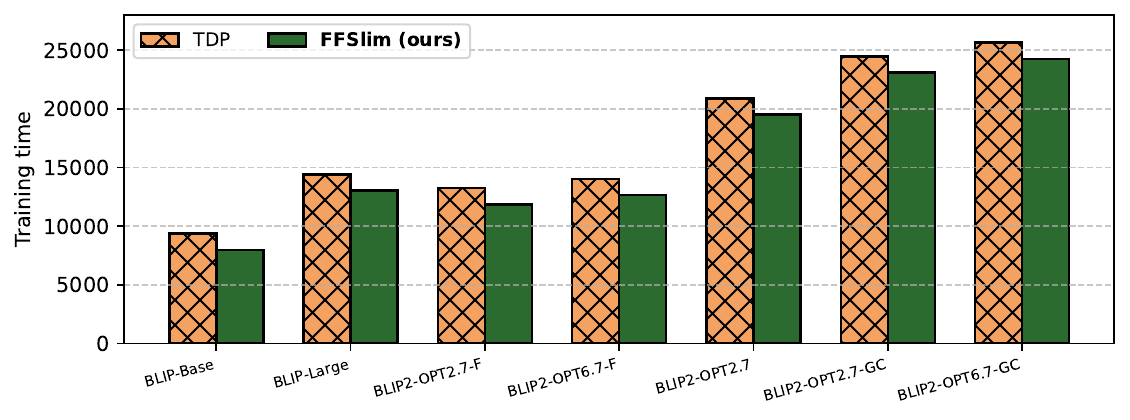}}
\caption{Training time of one epoch on MSCOCO under
different data formats across multi-modal
models. \footnotesize{Using \textit{FFSlim} to organize multi-modal datasets shortens training time, owing to its high data-loading efficiency.}}
\label{end-to-end_case_study}
\end{figure}

% 实验结果分析
Figure~\ref{end-to-end_case_study} illustrates the end-to-end epoch training time across all evaluated workloads. As depicted, \textit{FFSlim} achieves a substantial $5.36\%$--$14.18\%$ reduction in total epoch training time compared to \textit{TDP}. These macro-level performance gains inherently stem from the micro-level data access efficiency of \textit{FFSlim}. By significantly accelerating random sample access, \textit{FFSlim} successfully alleviates the severe I/O bottlenecks that typically stall training pipelines. Consequently, it seamlessly translates these low-level storage optimizations into tangible end-to-end training accelerations.

\subsection{Ablation Study}
% 实验负载介绍
We evaluate the data loading performance of \textit{FFSlim}, \textit{w/o Cache}, and \textit{w/o Format} under identical settings to quantify the contribution of each component. \textit{w/o Cache} removes the caching module from \textit{FFSlim}, while \textit{w/o Format} further disables the format module on top of \textit{w/o Cache}.

% 初步总体观察分析
Table~\ref{tab:ablation} shows the impact of each optimization in \textit{FFSlim} on data loading performance across seven datasets. Removing the cache leads to an average performance drop of 59.45\%, as it eliminates the acceleration gained from reusing other media–text pairs within a cached \textit{Media2Texts} object. Removing the format further degrades performance by 79.44\%, since the absence of a unified data format reintroduces the large number of small files and slows sample retrieval. Overall, both the cache and format components are essential to \textit{FFSlim}, with caching providing the dominant contribution to performance.

% 讨论泛化能力+局限性
\section{Discussion and Limitation}

\paragraph{Different dataset characteristics.}
The performance and overhead of different data formats are strongly influenced by fundamental dataset characteristics, including media count, media-to-text ratio, and media size.

As the number of media objects increases, \textit{Files} and \textit{TDP} produce an enormous number of media files, exacerbating the small-file problem and significantly degrading storage and retrieval performance. In contrast, \textit{FFRecord} and \textit{FFSlim} serialize the dataset into a single large file, inherently avoiding this issue and exhibiting only mild performance degradation. Media count also affects index memory usage: \textit{TDP} grows the fastest because it builds sample-level indexes and embeds text content directly into the index. \textit{FFRecord} grows more moderately, as it also indexes at the sample level. \textit{FFSlim} increases more slowly due to its coarser Media2Texts-level indexing. \textit{Files} incurs no memory overhead since it constructs indexes on the fly.

As the media-to-text ratio increases, redundancy in \textit{Files} and \textit{FFRecord} becomes increasingly significant. For a $1:N$ ratio, the additional $N-1$ duplicated media copies inflate storage consumption by nearly $N\times$, because media is much larger than text. In contrast, \textit{TDP} and \textit{FFSlim} introduce almost no redundancy, keeping storage consumption consistently low. The media-to-text ratio also shapes caching behavior. Higher ratios cause duplicated media to be accessed repeatedly within an epoch, raising cache value, improving hit rates, and boosting data loading performance for \textit{TDP} and \textit{FFSlim}. By design, \textit{Files} and \textit{FFRecord} remain cache-unfriendly and gain no benefit across all ratios. When the ratio is $1:1$, the dataset effectively degenerates into a uni-modal setting. No media duplication arises, and caching offers no improvement because every media item is accessed only once per epoch. In this rare but extreme case, the dominant challenge is again the small-file problem: \textit{FFRecord} and \textit{FFSlim} retain good performance, whereas \textit{Files} and \textit{TDP} degrade sharply.

As media size increases, the small-file overhead in \textit{TDP} becomes less significant, and its read performance gradually approaches that of \textit{FFSlim}. \textit{Files}, however, continues to suffer from the large number of text-generated small files regardless of media size.

\begin{table}[t]
\centering
\caption{Ablation study of data loading performance. \footnotesize{Both the cache and the data format contribute substantially to data-loading performance; removing either component leads to a significant performance drop.}}
\label{tab:ablation}
\footnotesize %\tiny %\small
% \resizebox{\columnwidth}{!}{%
\begin{tabular}{lccc} 
\toprule
\textbf{Workload} & \textbf{FFSLim} & \textbf{w/o Cache} & \textbf{w/o Format} \\
\midrule
Flickr30k & 567.45   & 256.61 & 116.50  \\
F30-E     & 565.55   & 259.63 & 116.27  \\
MSCOCO    & 541.88   & 255.95 & 107.87  \\
VQA2      & 629.46   & 277.45 & 102.69  \\
GQA       & 1,542.79 & 373.49 & 115.72  \\
MSRVTT    & 394.82   & 111.59 & 71.76  \\
Clotho    & 96.28    & 47.12  & 39.40  \\
\bottomrule
\end{tabular}
% }
\end{table}
\paragraph{Multi-modal dedicated caching framework.}
At present, FFSlim relies entirely on the system’s default page cache to accelerate repeated media access. However, the page cache uses an LRU-style replacement policy designed for workloads with strong temporal locality—an assumption that does not align well with the highly irregular and modality-dependent access patterns of multi-modal training. In addition, page caches are maintained independently on each compute node, resulting in fragmented cache capacity and unnecessary duplication of cached media across nodes. As future work, we plan to design a caching framework tailored specifically to multi-modal training. Such a system would centrally manage cached data and apply replacement strategies informed by modality-aware access patterns, thereby improving cache utilization under constrained memory budgets.

\section{Related Work}
\label{sec6}
\subsection{Data Format Optimizations}
To bypass underlying file system bottlenecks, modern datasets are typically serialized along two distinct trajectories. The first trajectory adopts pseudo-randomness (i.e., local shuffling) to maximize streaming bandwidth. Formats such as \textit{TFRecord}~\cite{tfrecord}, \textit{WebDataset}~\cite{webdataset}, and \textit{MDS}~\cite{mosaicml2022streaming} perform coarse-grained inter-shard shuffling and stream data sequentially. While this approach yields high throughput, sacrificing strict global randomness can compromise model convergence.
The second trajectory preserves exact \textit{sample-level point queries} to guarantee training accuracy. Within this paradigm, \textit{TDP}~\cite{alayrac2022flamingo} mitigates metadata overhead by caching all text data in memory and loading media objects on demand, deliberately trading excessive memory consumption for improved I/O performance. To address the small-file penalty, \textit{FFRecord}~\cite{ffrecord} packs independent media--text pairs into contiguous chunks. This design converts massive small-file random reads into intra-file random accesses within a single large file, entirely bypassing the severe overhead of frequent file open and close operations. 

However, existing formats share a fundamental structural flaw: by flattening datasets into isolated media--text pairs, they ignore the inherent one-to-many \textit{Media2Texts} relationship, inevitably causing severe I/O inefficiencies and prohibitive overhead. To address this, \textit{FFSlim} explicitly embraces these structural characteristics by fundamentally redesigning the storage layout and retrieval granularity. Coupled with dataset-aware adaptive indexing, \textit{FFSlim} delivers extreme I/O performance while strictly maintaining a minimal resource footprint.

\subsection{Training I/O Bottleneck Optimizations}
Beyond data format optimizations, the systems community has extensively explored deep learning (DL) I/O acceleration across three orthogonal dimensions: file system design, prefetching mechanisms, and cache management.

\textbf{File System Optimizations.} To prevent massive influxes of small multi-modal files from overwhelming metadata services, recent distributed file systems fundamentally redesign metadata architectures. Key trajectories include decoupling directory semantics~\cite{mantle2025}, scaling capacity via key-value or serverless backends~\cite{zhu2025fdbkeeper}, and mitigating synchronization bottlenecks~\cite{wang2025origami}. These foundational techniques culminate in specialized AI parallel file systems (\textit{3FS}~\cite{3fs2024}) designed to deliver extreme raw throughput.

\textbf{Prefetching Optimizations.} To hide storage latency, popular DL frameworks (PyTorch~\cite{paszke2019pytorch}), alongside numerous storage studies~\cite{ImPACT}, utilize multithreading and asynchronous I/O to overlap prefetching with computation. By proactively staging training data into memory buffers, these systems effectively mask I/O wait times and prevent GPU starvation. 

\textbf{Cache Management Policies.} Caching is crucial for mitigating I/O stalls during ML training, with policies evolving from general heuristics to workload-aware algorithms. Traditional policies (LRU~\cite{lru}) often fail because mandatory epoch-level random reshuffling actively destroys data locality, leading to severe cache thrashing. Consequently, DL-specific caches have emerged: \textit{Quiver}~\cite{kumar2020quiver} evicts data based on expected training benefits, while \textit{CoorDL}~\cite{coordl} relies on static data pinning to avoid thrashing under highly random access patterns. More recently, caches have been tailored for complex workloads, such as prioritizing sample importance in importance-sampling training (\textit{ImPACT}~\cite{ImPACT}) or leveraging topological dependencies in Graph Neural Networks (\textit{GNNLab}~\cite{GNNLab}).

Importantly, these infrastructural optimizations are strictly orthogonal to \textit{FFSlim}. While they accelerate I/O at the backend storage or runtime pipeline level, \textit{FFSlim} focuses exclusively on the data layout format, allowing it to be seamlessly integrated with and further amplify the benefits of these existing systems.

\section{Conclusion}
Data I/O has emerged as a severe bottleneck in multi-modal training, largely because existing formats ignore the inherent characteristics of multi-modal datasets. In this paper, we propose \textit{FFSlim}. By fundamentally redesigning the data layout, management and retrieval granularity, and dataset-aware adaptive indexing structures, \textit{FFSlim} systematically resolves four critical limitations of existing formats: storage redundancy, massive small-file I/O, cache-unfriendly layouts, and prohibitive index overhead. Consequently, it simultaneously achieves extreme I/O performance and ultra-low system cost. Our comprehensive evaluations show that \textit{FFSlim} outperforms existing formats by $2.07\times$ and $8.26\times$ in read and write throughputs, respectively, with a negligible resource footprint. Ultimately, these structural optimizations yield a $5.36\%$--$14.18\%$ reduction in end-to-end epoch training time, positioning \textit{FFSlim} as a highly scalable default format for future multi-modal AI infrastructure.
%-------------------------------------------------------------------------------
\bibliographystyle{IEEEtran}
% \bibliography{citation.bib}

\begin{IEEEbiography}[{\includegraphics[width=1in,height=1.25in,clip,keepaspectratio]{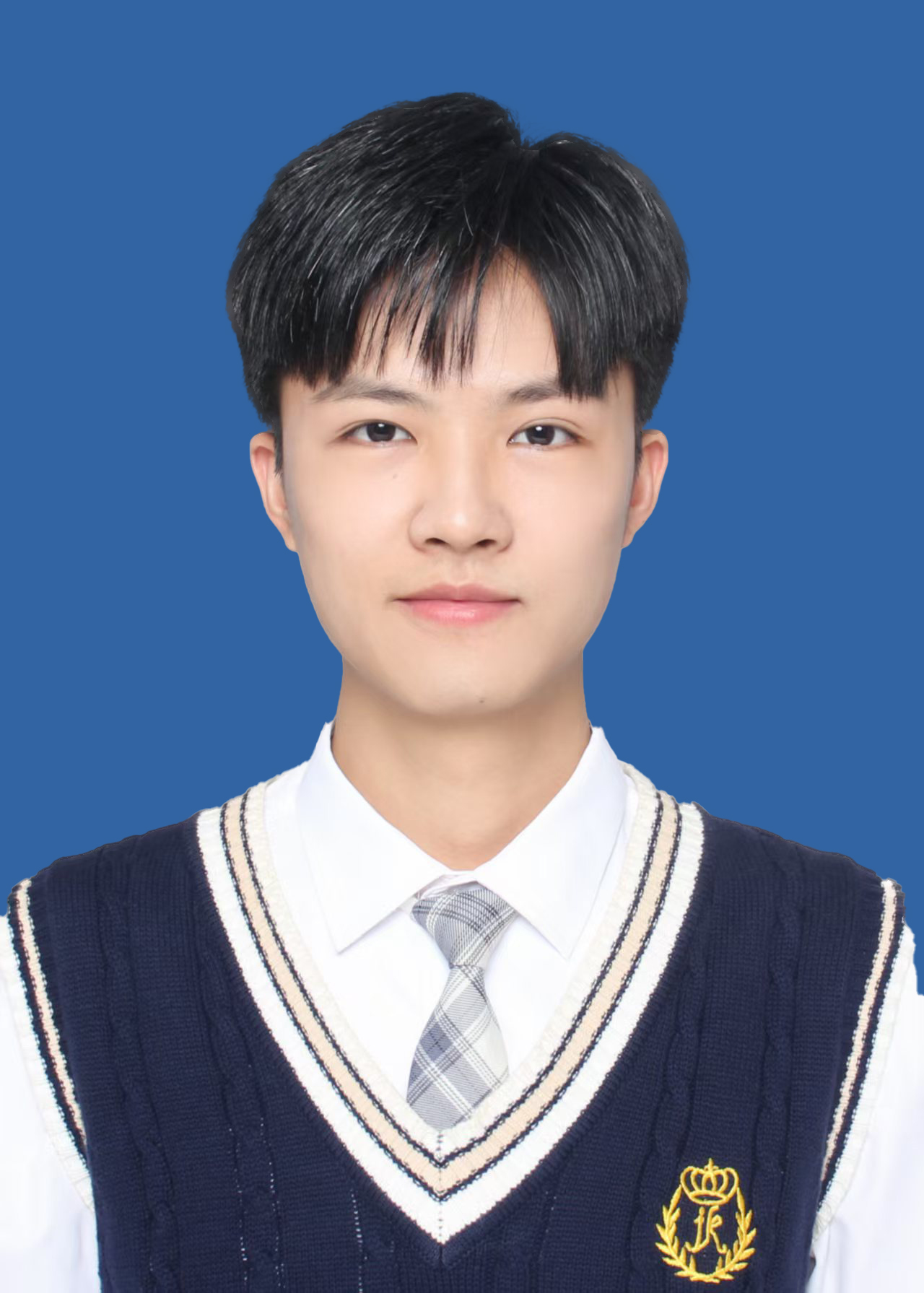}}]{Long Yang}
is currently working toward the Ph.D. degree in a joint program between East China Normal University and Shanghai Innovation Institute, China. His research interests include AI training optimization, storage system optimization, and AI infrastructure.
\end{IEEEbiography}

\begin{IEEEbiography}[{\includegraphics[width=1in,height=1.25in,clip,keepaspectratio]{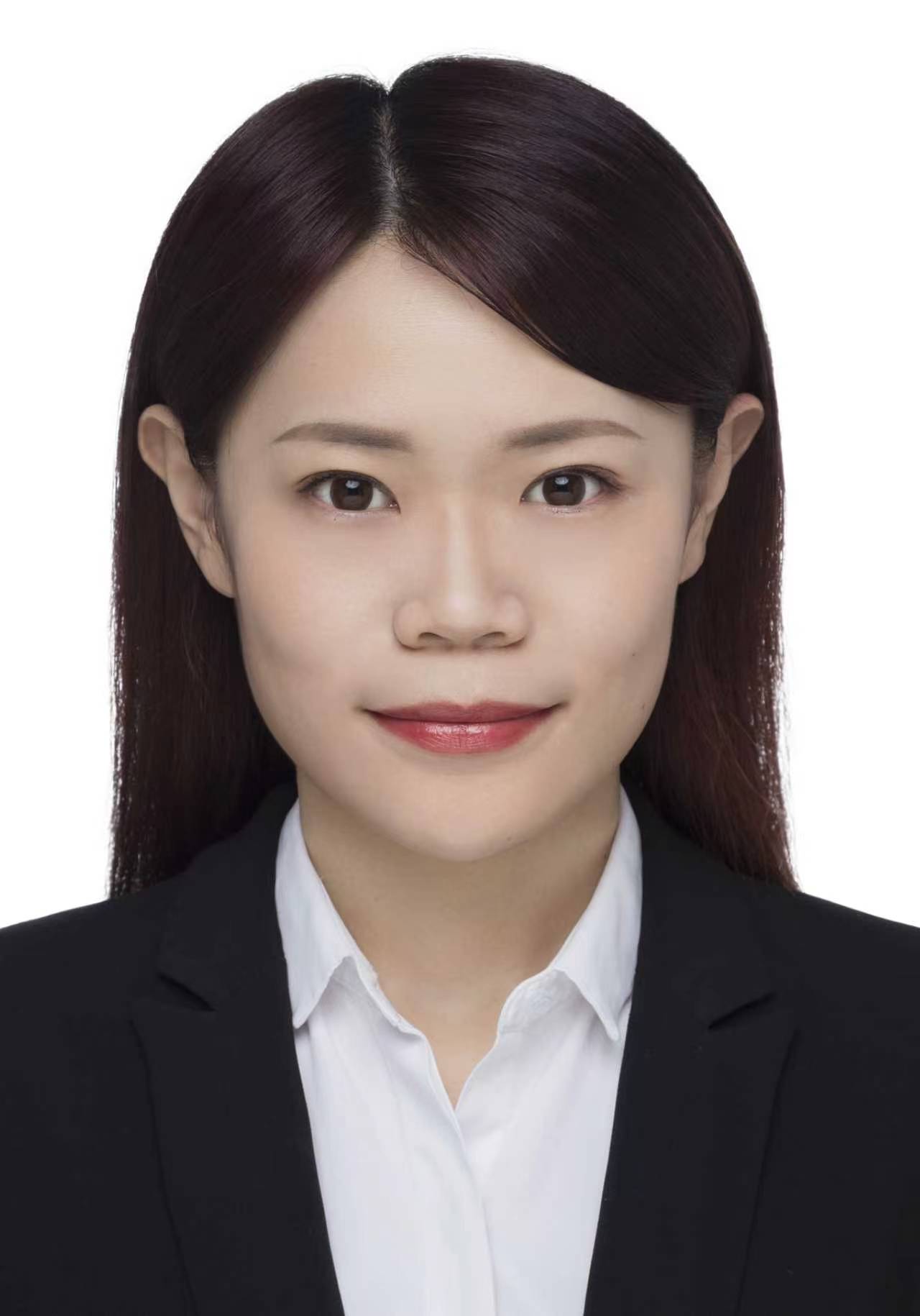}}]{Yu Mao}
received the Ph.D. degree in computer science from City University of Hong Kong, Hong Kong. Yu is currently a Research Scientist with ByteDance, San Jose, CA, USA. Research interests include neural and learned data compression, system efficiency at scale, and hardware-aware machine learning acceleration.
\end{IEEEbiography}

\begin{IEEEbiography}[{\includegraphics[width=1in,height=1.25in,clip,keepaspectratio]{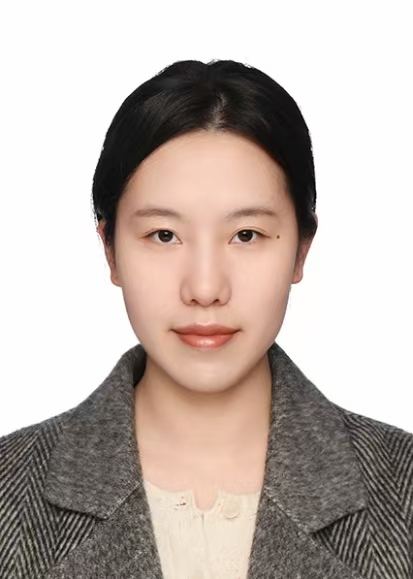}}]{Yuchen Shao}
 is currently a third-year Ph.D. student in the Software Engineering Institute at East China Normal University and Shanghai Innovation Institute, co-advised by Prof. Chengcheng Wan and Prof. Ting Su. Her research interests include software engineering for AI and LLM security.
\end{IEEEbiography}

\begin{IEEEbiography}[{\includegraphics[width=1in,height=1.25in,clip,keepaspectratio]{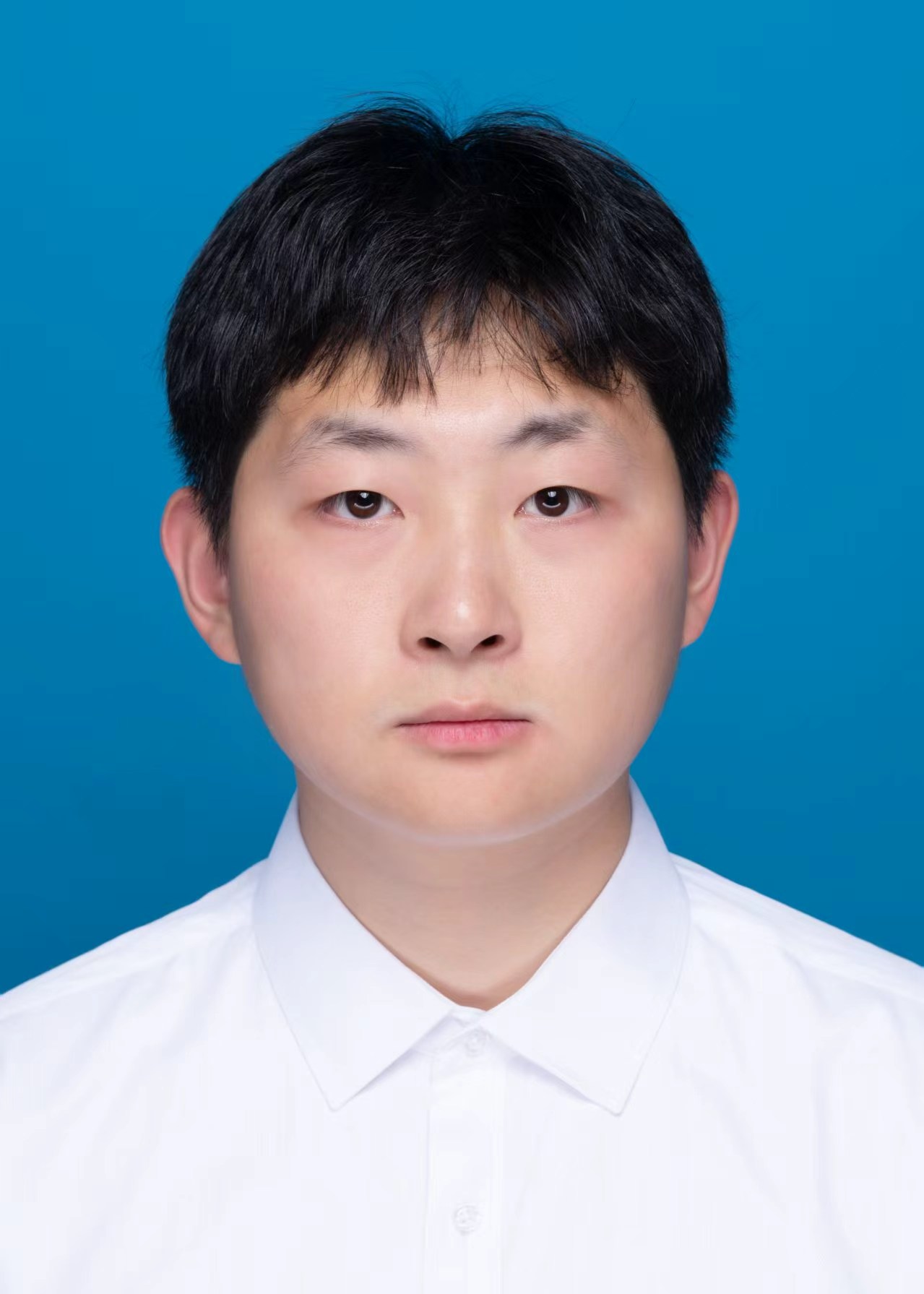}}]{Yumiao Zhao}
 is currently working toward the Ph.D. degree in a joint program between East China Normal University and the Shenzhen Loop Area Institute, China. His research interests include AI infrastructure, storage systems, and computational networking.
\end{IEEEbiography}

\begin{IEEEbiography}[{\includegraphics[width=1in,height=1.25in,clip,keepaspectratio]{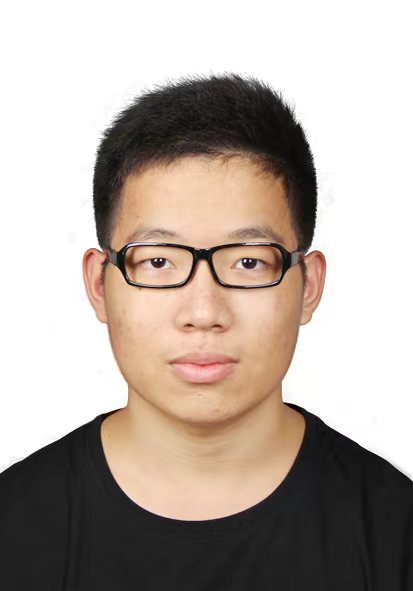}}]{Yaqi Li}
received the B.E. degree from Jiangnan University, Wuxi, China, in 2023. He is currently working toward the Ph.D.degree in a joint program between East China Normal University and Shanghai Innovation Institute,China.His research interests include AI inference systems, storage systems, systems for AI, and AI infrastructure.
\end{IEEEbiography}

\begin{IEEEbiography}[{\includegraphics[width=1in,height=1.25in,clip,keepaspectratio]{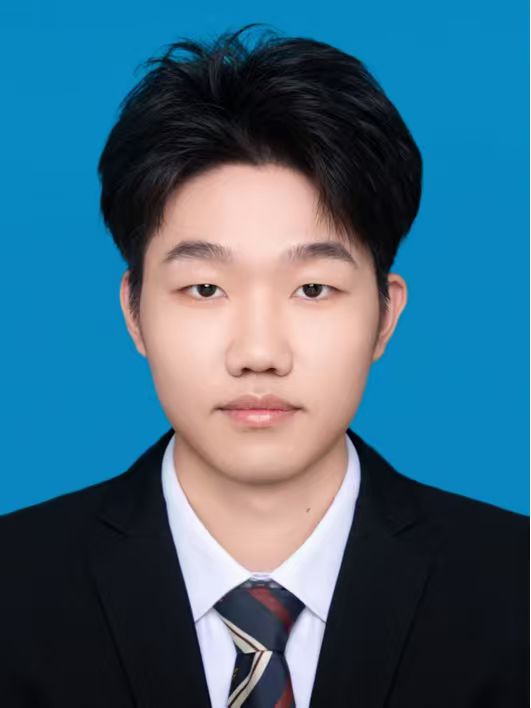}}]{Xuan Liu}
is currently pursuing a graduate degree at East China Normal University. His research interests include storage system optimization, data processing, and query engine design and optimization.
\end{IEEEbiography}

\begin{IEEEbiography}[{\includegraphics[width=1in,height=1.25in,clip,keepaspectratio]{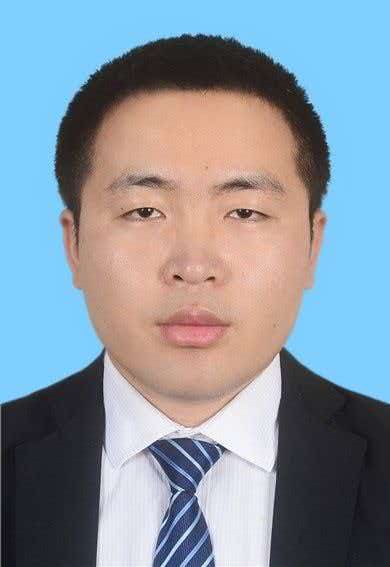}}]{Xiaolong Shen}
received the Ph.D. degree from the College of Computer Science and Technology, National University of Defense Technology, Changsha, China. He is a Senior Engineer with the Central Research Institute, Huawei Technologies Company Ltd., Shenzhen, China. His main research interests include advanced computing and storage, AI applications, computer architecture, and high-performance computing.
\end{IEEEbiography}

\begin{IEEEbiography}[{\includegraphics[width=1in,height=1.25in,clip,keepaspectratio]{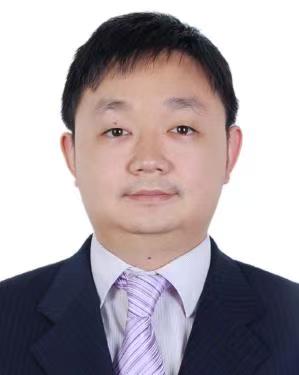}}]{Tao Yu}
received M.S.in Pattern Recognition and Intelligent Systems from Huazhong University of Science and Technology, is a Senior Engineer at Huawei's Central Research Institute in Shanghai. His interests span advanced computing and storage, AI data storage, training and inference frameworks, and high-performance computing.
\end{IEEEbiography}

\begin{IEEEbiography}[{\includegraphics[width=1in,height=1.25in,clip,keepaspectratio]{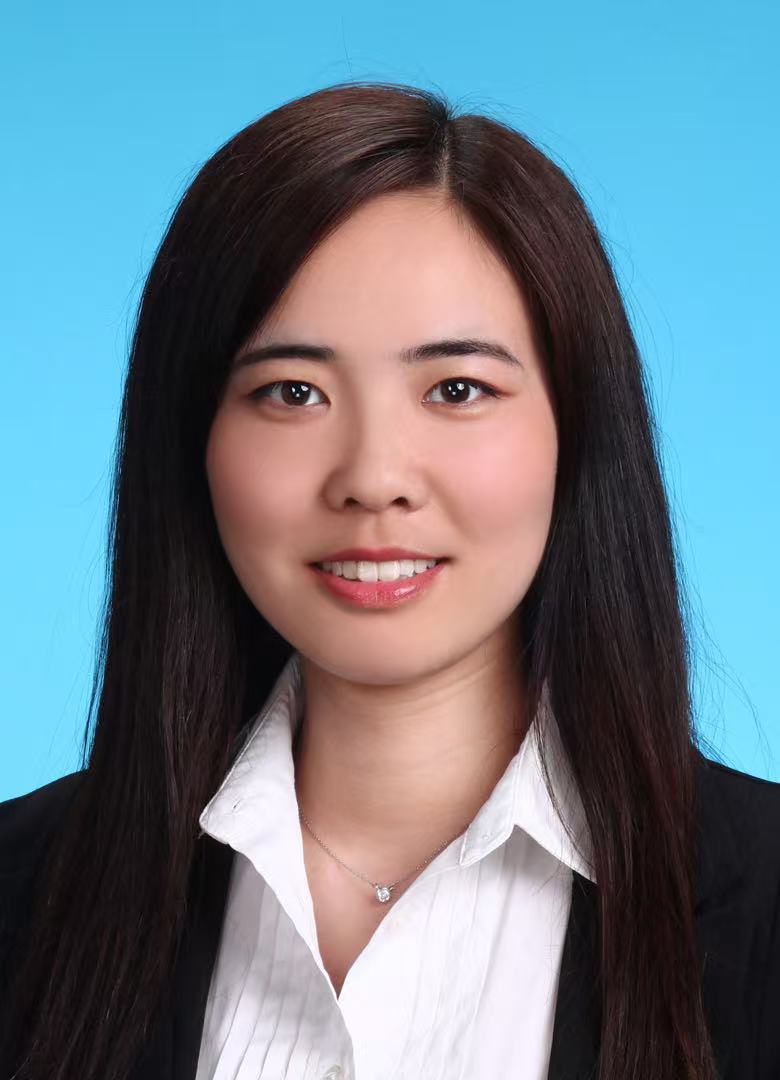}}]{Gezi Li}
is a technical expert at the Central Research Institute of Huawei Technologies Co., Ltd.
She received her Ph.D. from the Shanghai Institute of Microsystem and Information Technology, Chinese Academy of Sciences. Her main research focuses on AI data storage, including KV cache for LLM inference (using retrieval instead of recomputation), agent memory, memory/storage systems in AI infrastructure, AI SSDs, etc.
\end{IEEEbiography}

\begin{IEEEbiography}[{\includegraphics[width=1in,height=1.25in,clip,keepaspectratio]{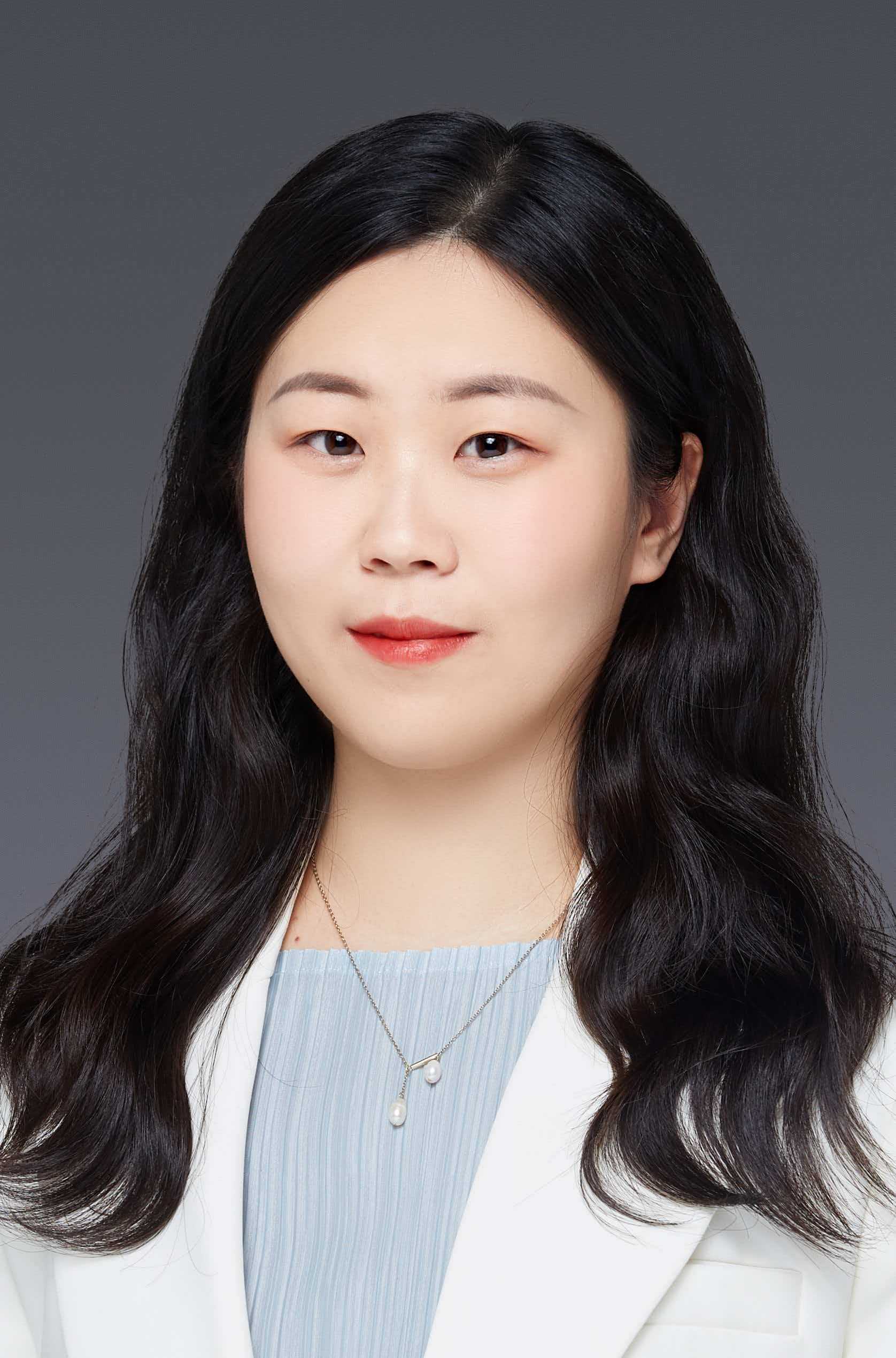}}]{Jing Wang}
is an assistant professor at East China Normal University. She received her Ph.D. from Shanghai Jiao Tong University. Her research focuses on high-performance memory and storage systems.
\end{IEEEbiography}

\begin{IEEEbiography}[{\includegraphics[width=1in,height=1.25in,clip,keepaspectratio]{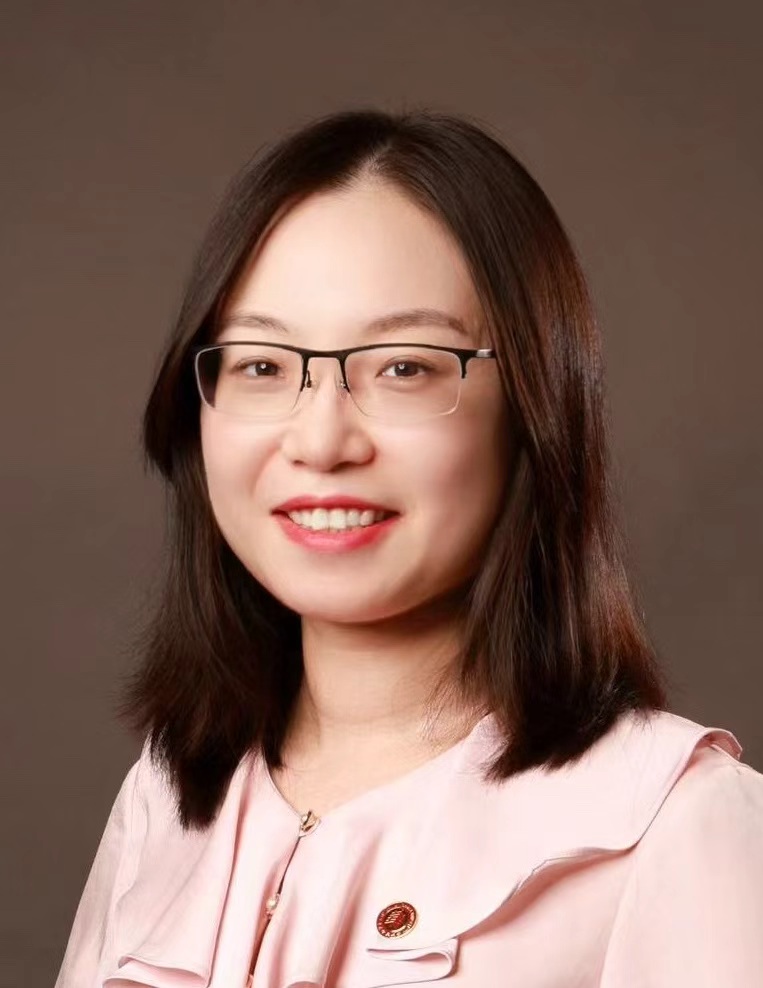}}]{Chengcheng Wan}
received the B.Eng. degree in Computer Science and Technology from
Shanghai Jiao Tong University in 2017 and, Ph.D. degree in Computer Science from the University of Chicago in 2022. She is currently an associate professor at East China Normal University. Her research interests including ML system, SE for AI and AI for SE.
\end{IEEEbiography}

\begin{IEEEbiography}[{\includegraphics[width=1in,height=1.25in,clip,keepaspectratio]{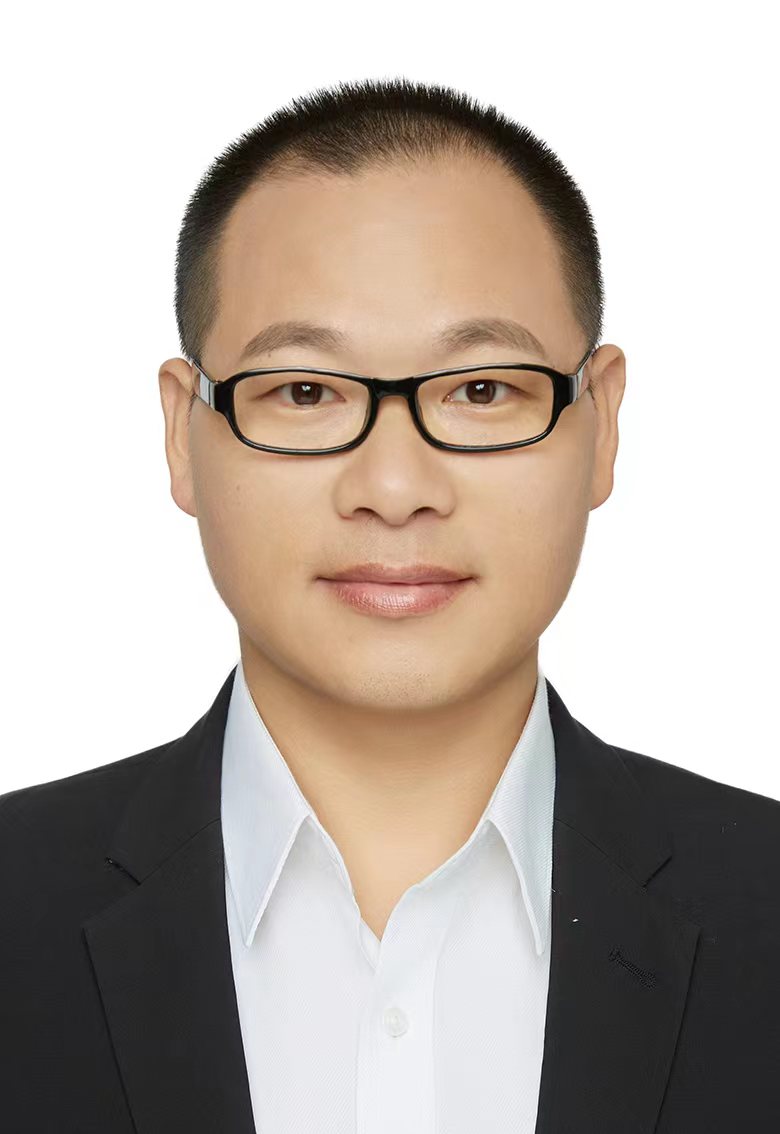}}]{Liang Shi}
received the B.S. degree in Computer Science from Xi'an University of Post \& Telecommunication, Xi'an, Shanxi, China, in July, 2008, Ph.D. degree from University of Science and Technology of China, Hefei, China, in June, 2013. Since 2013, he has been an associate professor with the College of Computer Science, Chongqing University, Chongqing, China. Since 2018, he has been a full professor with the School of Computer Science and Technology, East China Normal University.
His research interests include flash memory, embedded systems, and emerging non-volatile memory technology.
\end{IEEEbiography}
\end{document}

%% file: micros.tex
\newcommand{\eg}{\hbox{\emph{e.g.}}\xspace}

\newcommand{\cmark}{\textcolor[rgb]{0,0.5,0}{\checkmark}} 
\newcommand{\xmark}{\textcolor{red}{\ding{55}}} 

\definecolor{wsjfiles}{RGB}{199, 199, 199}    
\definecolor{wsjtdp}{RGB}{174, 132, 126}   
\definecolor{wsjffrecord}{RGB}{242, 157, 100}     
\definecolor{wsjffslim}{RGB}{107, 142, 208}